\documentclass[twocolumn]{aastex631}
\usepackage{apjfonts}
\usepackage{nccmath}
\usepackage{stackengine}
\usepackage{natbib}
\usepackage[htt]{hyphenat}
\usepackage{blindtext}
\usepackage{rotating}
\usepackage{graphicx}
\usepackage{graphics,graphicx,float,import,enumitem,booktabs}

\newcommand{\symuncert}[2]{\ensuremath{#1 \pm #2}}
\newcommand{\asymuncert}[3]{\ensuremath{#1^{+#2}_{-#3}}}

\newcommand\siii{[\ion{S}{3}]}

\shorttitle{CECILIA: Ultra-deep rest-optical spectra of continuum-faint galaxies at $z$$\sim$2.5}
\shortauthors{Raptis et al.}

\begin{document}

\title{CECILIA: Ultra-Deep Rest-Optical Spectra of Faint Galaxies at Cosmic Noon}

\author[0009-0008-2226-5241]{Menelaos Raptis}
\affiliation{Department of Physics and Astronomy, Franklin \& Marshall College, 637 College Avenue, Lancaster, PA 17603, USA}
 
\author[0000-0002-6967-7322]{Ryan F. Trainor}
\affiliation{Department of Physics and Astronomy, Franklin \& Marshall College, 637 College Avenue, Lancaster, PA 17603, USA}
\affiliation{William H. Miller III Department of Physics and Astronomy, Johns Hopkins University, Baltimore, MD 21218, USA}

\author[0000-0001-6369-1636]{Allison L. Strom}
\affiliation{Center for Interdisciplinary Exploration and Research in Astrophysics (CIERA), Northwestern University, 1800 Sherman Ave., Evanston, IL, 60201, USA}
\affiliation{Department of Physics and Astronomy, Northwestern University, 2145 Sheridan Road, Evanston, IL 60208, USA}

\author[0000-0002-8459-5413]{Gwen C. Rudie}
\affiliation{The Observatories of the Carnegie Institution for Science, 813 Santa Barbara Street, Pasadena, CA 91101, USA}

\author[0000-0002-0361-8223]{Noah S. J. Rogers}
\affiliation{Center for Interdisciplinary Exploration and Research in Astrophysics (CIERA), Northwestern University, 1800 Sherman Ave., Evanston, IL, 60201, USA}

\author[0000-0002-4834-7260]{Charles C. Steidel}
\affiliation{Cahill Center for Astronomy and Astrophysics, California Institute of Technology, MS 249-17, Pasadena, CA 91125, USA}

\author[0000-0003-0695-4414]{Michael V. Maseda}
\affiliation{Department of Astronomy, University of Wisconsin-Madison, 475 N. Charter Street, Madison, WI 53706, USA}

\author[0000-0002-6034-082X]{Caroline von Raesfeld}
\affiliation{Center for Interdisciplinary Exploration and Research in Astrophysics (CIERA), Northwestern University, 1800 Sherman Ave., Evanston, IL, 60201, USA}
\affiliation{Department of Physics and Astronomy, Northwestern University, 2145 Sheridan Road, Evanston, IL 60208, USA}

\author[0000-0003-2385-9240]{Nathalie A. Korhonen Cuestas}
\affiliation{Center for Interdisciplinary Exploration and Research in Astrophysics (CIERA), Northwestern University, 1800 Sherman Ave., Evanston, IL, 60201, USA}
\affiliation{Department of Physics and Astronomy, Northwestern University, 2145 Sheridan Road, Evanston, IL 60208, USA}

\defcitealias{Trainor2016}{T16}
\defcitealias{stro2023}{S23}
\defcitealias{2024ApJ...964L..12R}{R24}
\defcitealias{stro2017}{S17}

\begin{abstract}
Intrinsically faint galaxies at $z\sim2-3$ offer critical insights into early galaxy formation, tracing low-metallicity, low-mass systems during Cosmic Noon and serving as analogs to reionization-era galaxies. We present ultra-deep JWST/NIRSpec spectroscopy of nine low-luminosity galaxies ($-17 \lesssim M_{\rm UV} \lesssim -20$, $M_\star \lesssim 10^9\,M_\odot$) at $z\sim2.5$ from the CECILIA program, with $\sim$29.5 hr in G235M/F170LP and 1 hr in G395M/F290LP. Our sample includes four LAEs, three rest-UV color–selected galaxies, and two serendipitous detections — providing the most sensitive rest-optical spectra of individual faint galaxies at this epoch to date. Balmer-line measurements reveal low SFRs ($0.63 < \mathrm{SFR}/(M_\odot\,\mathrm{yr}^{-1}) < 5.43$) and a broad range of dust reddening ($0 < E(B-V) < 1$), with SFRs systematically below those of continuum-selected galaxies. Electron densities are low ($n_e \lesssim 200$cm$^{-3}$), and emission-line diagnostics indicate low [NII]/H$\alpha$, high [OIII]/H$\beta$, suggesting metallicities $12+\log({\rm O/H})\lesssim8.0$. We also present the first O1-BPT constraints in such faint high-redshift galaxies. Notably, two galaxies show low [OIII]/H$\beta$ despite high Ly$\alpha$ EWs and very low [NII]/H$\alpha$, consistent with the predicted turnover in this ratio at very low metallicities — highlighting the need for complementary diagnostics (e.g., N2, O32) to identify metal-poor systems. Direct $T_e$-based abundances and expanded samples are needed to further trace metallicity and ionization trends in low-mass galaxies.

\end{abstract}

\section{Introduction}
\label{Introduction}
One of the central goals of extragalactic astrophysics is to understand how galaxies evolve across cosmic time. What drives the differences in their ionization states, chemical abundances, and star formation histories? How do galaxies at $z\sim2-3$ — often referred to as "Cosmic Noon," the peak of star formation and gas accretion \citep{2014ARA&A..52..415M} — compare to those in the local universe? Can they serve as analogs for galaxies at $z \gtrsim 6$ during the Epoch of Reionization (EoR)? Answering these questions requires detailed studies of their optical nebular emission lines, which provide crucial diagnostics for ionization conditions, chemical compositions, and star formation activity.


Recombination lines of hydrogen provide constraints on star formation rates (SFRs) and reddening due to dust, while collisionally excited forbidden transitions of elements like oxygen, nitrogen, and sulfur are sensitive to nebular properties such as the ionization state and gas-phase enrichment, in addition to discriminating between star-forming galaxies and active galactic nuclei \citep[AGNs;][]{Veilleux1987, 2004astro.ph..6220B, Kewley2006}.
Ground-based facilities such as Keck/MOSFIRE, VLT/X-Shooter, and VLT/KMOS have provided initial constraints on galaxy populations at Cosmic Noon (e.g., \citealt{stro2017}, hereafter \citetalias{stro2017}; \citealt{Shapley2019,sand2020, Sanders2021, Strom2022}), but their reach is fundamentally limited by sky background, atmospheric absorption, and a bias toward relatively massive and luminous galaxies. The launch of JWST has significantly extended the study of nebular emission in distant galaxies, enabling rest-frame optical spectroscopy out to Cosmic Noon and the EoR. Several JWST programs (e.g., CECILIA, AURORA, EXCELS, MARTA; \citealt{stro2023}, hereafter \citetalias{stro2023}; \citealt{Shapley21, Carnall2024, Cataldi2025}) are now focused on characterizing the detailed chemical properties at $z > 2$, allowing for unprecedented studies of typical ($L \sim L_{\ast}$) galaxies in the early universe.

 Despite the progress enabled by JWST, studies of sub-$L_{\ast}$ galaxies at Cosmic Noon remain challenging due to their lower luminosities, making them harder to detect and characterize. However, these low-mass, continuum-faint galaxies are particularly interesting because they provide crucial insight into how key physical properties, such as metallicity and ionization, evolve at lower stellar masses and SFRs. Moreover, they appear to be better analogs for the galaxies that dominated the EoR, as they are less evolved and share similar masses and ionization conditions (\citealt{Trainor2016}, hereafter \citetalias{Trainor2016}; \citealt{Tang2019, sand2020}). At fixed M$_{\text{UV}}$, the fraction of strong Ly$\alpha$ emitters (LAEs) increases with redshift out to $z \sim 6$ \citep[]{Stark2010, Stark2011}. Beyond $z \sim 6$, the abundance of LAEs diminishes due to the absorption and scattering of Ly$\alpha$ photons by neutral hydrogen in the intergalactic medium \citep[IGM;][]{Pentericci2011, Schenker2012}. However, the intrinsic Ly$\alpha$ emission of galaxies, excluding attenuation from the IGM and circumgalactic medium (CGM), likely continues to rise towards earlier cosmic epochs. Therefore, low-mass, continuum-faint galaxies --- particularly those with strong Ly$\alpha$ emission --- present an opportunity to gain insight into the physical conditions and star formation rates at even higher redshifts.

LAEs have been studied across a wide range of cosmic epochs through various spectroscopic surveys, from $z \sim 0$ \citep[e.g., LARS;][]{Ostlin2014} to $z > 6$ \citep[e.g.,][]{Hu2017, Runnholm25, Willott25, Messa25}. However, the properties of faint LAEs have remained challenging to constrain at all redshifts, and nebular spectroscopy of intrinsically faint LAEs is still very limited. Some of the earliest measurements of $L \ll L_\ast$ LAEs came from Ly$\alpha$-selected samples in the Keck Baryonic Structure Survey (KBSS-Ly$\alpha$; \citealt{trai2015}; \citetalias{Trainor2016}), which used stacked spectra to examine their average properties. Additional insights came from \citet{Gburek2019}, who studied nebular emission in gravitationally lensed LAEs.

\citet{Maseda2023} analyzed ultra-faint galaxies (M$_{\text{UV}} \sim -15$; 0.01 $L_*$) at $z = 2.9-6.7$ with JWST/NIRSpec, finding a relationship between strong-line ratios and Ly$\alpha$ equivalent widths (EWs), suggesting a connection between gas-phase metallicity and Ly$\alpha$ emission. This trend appears to be a continuation of similar correlations measured for $L \gtrsim 0.1 L_*$ galaxies by \citetalias{Trainor2016} and \citet{Trainor2019} who found a strong association between Ly$\alpha$ EW and nebular emission-line ratios such as [O~III]/H$\beta$. While stacking techniques remain a common approach for studying these faint sources, individual detections are essential for understanding the diversity of their physical properties and ionization conditions.


In this paper, we investigate the rest-frame optical properties of continuum-faint galaxies at $z\sim2-3$, focusing on both their average characteristics and the diversity among individual sources. Specifically, we analyze nine of the faintest galaxies in the CECILIA sample \mbox{\citepalias{stro2023}}, including faint \mbox{Ly$\alpha$}-selected galaxies and other continuum-faint sources expected to have low metallicities and high ionization states. 
We construct emission-line diagnostic diagrams \citep{Baldwin1981, Veilleux1987}, derive H$\alpha$-based SFRs, and explore chemical enrichment using multiple diagnostics for the first time at $z > 2$ in a sample of extremely low-mass, continuum-faint galaxies. These analyses are performed using a combination of individual and stacked emission-line measurements, allowing us to constrain the physical conditions of these galaxies and place them in context with both local and high-redshift populations.

This manuscript is organized as follows: In \S2, we describe the JWST observations, data selection and reduction, and spectral extraction. In \S3, we present diagnostic diagrams and analyze the dust reddening, SFRs, and electron densities of the selected galaxies. In \S4, we compare our measurements to photoionization model grids and other samples of low-metallicity galaxies. Finally, we conclude in \S5.

In this work, we assume a $\Lambda$CDM cosmology with $H_0$$=$70 km s$^{-1}$ Mpc$^{-1}$, $\Omega_m$$=$0.3, and $\Omega_\Lambda$$=$0.7. We adopt the solar abundance values from \citet{aspl2021}: 12+log(O/H)$_\odot = 8.69\pm0.04$ dex. Throughout the text, we refer to emission lines using their air wavelengths.

\section{Data}
\subsection{
Observations \& Target Selection}

The galaxies in this study are drawn from the CECILIA survey (\citetalias{stro2023}; \citealt{2024ApJ...964L..12R}, hereafter \citetalias{2024ApJ...964L..12R}). CECILIA is a JWST NIRSpec/MSA program to conduct ultra-deep observations of galaxies selected from the KBSS (\citealt{rudi2012,stei2014}; \citetalias{stro2017}) and KBSS-Ly$\alpha$ (\citealt{trai2015}, \citetalias{Trainor2016}) surveys to cover a broad range of galaxy properties in the Q2343 field. Targeting multiple electron temperature-sensitive auroral lines, CECILIA aims to enhance our understanding of the physical conditions and chemical evolution of high-redshift galaxies by analyzing the rest-optical and rest-NIR spectra ($\lambda_{\text{rest}} \approx 0.5-1.2\,\mu$m) of star-forming galaxies at redshifts $z\approx 2-3$.

The observational strategies, data reduction, and sample selection for the CECILIA program are described in detail by \citetalias{stro2023}, with further analysis strategies described by N. Rogers et al. (in prep.). Briefly, the CECILIA observations were conducted with JWST/NIRSpec in the multi-object spectroscopy (MOS) mode for a single mask targeting 34 galaxies, including ultra-deep (29.5~hr) observations in G235M/F170LP and moderately deep (1.1~hr) observations in G395M/F290LP. The data were reduced using the \textsc{calwebb} v1.15.1 pipeline with \textsc{crds\textunderscore context} 1251.pmap \citep{calwebb_v1.12.5_2023}. We also use \textsc{grizli} v1.11.11 \citep{grizli}, \textsc{NSClean} \citep{raus2023}, and \textsc{msaexp} v0.8.5 \citep{msaexp} for additional steps in the reduction. Lastly, we use custom software to estimate and remove a global background model from the full CECILIA sample, as well as to correct for the 2D barshadows produced by the NIRSpec MSA as described by \citetalias{stro2023} and N. Rogers et al. (in prep.). The resulting data products are 2D spectrograms that have been rectified and drizzled onto a common grid, with each column having a fixed observed wavelength and each row having a fixed spatial position along the slit. An example 2D spectrogram is shown in the top panel of Figure~\ref{fig:1D spectrum}.

This study analyzes a subsample of 9 continuum-faint galaxies included in the CECILIA program; individual properties of each source are given in Table~\ref{t:galaxies}. Four sources (NB2929, NB2089, NB2875, and NB2571) were selected as Ly$\alpha$ emitters (LAEs) through narrowband (NB) imaging as part of the KBSS-Ly$\alpha$ survey. KBSS-Ly$\alpha$ galaxies were prioritized for inclusion on the CECILIA Micro-shutter Assembly (MSA) mask on the basis of spectroscopic confirmation of their Ly$\alpha$ emission with Keck/LRIS \citep{oke1995,steidel2004} as well as previous detections of H$\alpha$ and/or [\ion{O}{3}] with Keck/MOSFIRE \citep{mclean2012}. The narrowband selection and rest-UV spectroscopy for these sources is described in detail in \citet{trai2015} and \citet{Trainor2025}, while the previous rest-optical spectroscopy is described by \citet{trai2015} and \citetalias{Trainor2016}.

Three additional sources (fBM40, fBM47, and fC23) were selected from the KBSS on the basis of their rest-UV continuum colors using the three-band ($U_n G \mathcal{R}$) criteria described by \citet{steidel2004}. While the primary KBSS galaxy selection imposes a rest-UV flux limit ($\mathcal{R}<25.5$), the three KBSS objects included here are part of an extended sample\footnote{The extension of the \citet{steidel2003,steidel2004} continuum-color selection to $\mathcal{R}=25.5-26.5$ is described briefly by \citet{reddy2009}.} that includes objects with lower continuum luminosities; we designate these objects as ``fLBGs'' for the remainder of this work, where the ``f'' denotes their faint rest-UV magnitudes. The fLBGs observed for CECILIA were selected from the full KBSS sample using the inclusion criteria described by \citetalias{stro2023}.

Finally, two objects (C31b and BX587b) were discovered as serendipitous sources over the course of reducing the NIRSpec/MSA spectra. Both objects were visually identified in the 2D spectrograms of the primary sources (C31 and BX587, respectively) on the basis of their H$\beta$,  [\ion{O}{3}] $\lambda\lambda$4959,5007, and H$\alpha$ emission. In both cases, the serendipitous objects have small spatial offsets (<1\farcs0) from their respective primary sources, but they have sufficiently large separations in redshift ($\gtrsim$2500$\,\mathrm{km}\,\mathrm{s}^{-1}$) that each is unlikely to be physically associated with its targeted primary. The inclusion of these two objects in this study of continuum-faint galaxies is motivated by the lack of visible continuum emission in the 2D spectrum of either serendipitous source, as well as the general similarity in their redshifts and rest-optical emission line properties to the LAEs and fLBGs described above.

As discussed by \citet{trai2015} the KBSS-Ly$\alpha$ LAEs have dynamical masses $M_\textrm{dyn}\sim 10^8-10^9\,\mathrm{M}_\odot$, consistent with stellar mass estimates of the LAE and fLBG samples from spectral energy distribution (SED) modeling (e.g., \citealt{korh2025}). Due to the lack of resolved imaging to distinguish the serendipitous sources from their more luminous primary targets, we do not calculate mass estimates for those two sources, although their velocity dispersions and sizes along the NIRSpec/MSA slits are similar to the other objects in the sample (Table~\ref{t:galaxies}). Further analysis of the stellar masses and  SEDs of the CECILIA continuum-faint sample will be presented in future work.

\subsection{Spectral Extraction}
\begin{deluxetable*}{lccccc}
\tablecaption{Galaxy Names and Properties \label{t:galaxies}}
\tablewidth{\textwidth}
\tablehead{
  \colhead{Galaxy Name} &
  \colhead{Redshift} &
  \colhead{Ly$\alpha$ EW (\AA)} &
  \colhead{M$_{\text{UV}}$} &
  \colhead{E(B-V)} &
  \colhead{SFR (M$_{\odot}$/year)} 
}
\startdata
Q2343-NB2929 & 2.546 & \asymuncert{23}{19}{10} & $-18.99$ & \symuncert{0.43}{0.21} & \asymuncert{1.26}{0.88}{0.35}\\
Q2343-NB2089 & 2.571 & \asymuncert{73}{130}{46} & $> -17.29$ & \symuncert{0.28}{0.11} & \asymuncert{1.54}{0.51}{0.30} \\
Q2343-NB2875 & 2.545 & \asymuncert{320}{580}{78} & $> -17.74$ & \symuncert{0.05}{0.07} & \asymuncert{0.86}{0.15}{0.11} \\
Q2343-NB2571 & 2.577 & \asymuncert{47}{93}{34} & $-17.21$ & \symuncert{0.07}{0.08} & \asymuncert{0.63}{0.14}{0.10}  \\
Q2343-fBM47  & 2.278 & \asymuncert{20}{5}{5} & $-20.65$ & \symuncert{0.19}{0.14} & \asymuncert{2.88}{1.32}{0.67} \\
Q2343-fBM40  & 2.147 & ... & $-20.43$ & \symuncert{0.19}{0.12} & \asymuncert{3.31}{1.17}{0.67} \\
Q2343-fC23   & 2.173 & \asymuncert{-9}{3}{3} & $-19.85$ & \symuncert{0.47}{0.29} & \asymuncert{3.94}{5.42}{1.34} \\
Q2343-C31b   & 3.022 & ... & ... & \symuncert{0.23}{0.18} & \asymuncert{1.49}{0.87}{0.38} \\
Q2343-BX587b & 2.775 & ... & ... & \symuncert{0.95}{0.23} & \asymuncert{5.43}{4.98}{1.61} \\
\enddata

\tablecomments{Names and properties of the galaxies included in this investigation. The last two entries of the table refer to the serendipitous sources. \( E(B-V) \) and SFRs are calculated using previous Keck/MOSFIRE measurements \citepalias{stro2017} for the CECILIA fLBGs.}
\end{deluxetable*}

The process of spectral extraction differed for targeted and serendipitous sources, as we describe in the following subsections.
\subsubsection{Targeted Sources}
\label{sec:main_sources}
From the reduced 2D spectrograms, we extract the corresponding 1D spectra. This extraction process requires the careful creation of spatial profiles to accurately determine the position of each source in the 2D spectrograms. The determination of the best extraction width for every source is conducted through the following steps.

\begin{figure*}[!t]
   \centering
   \hspace*{-22mm}
    \includegraphics[width=1.3\textwidth]{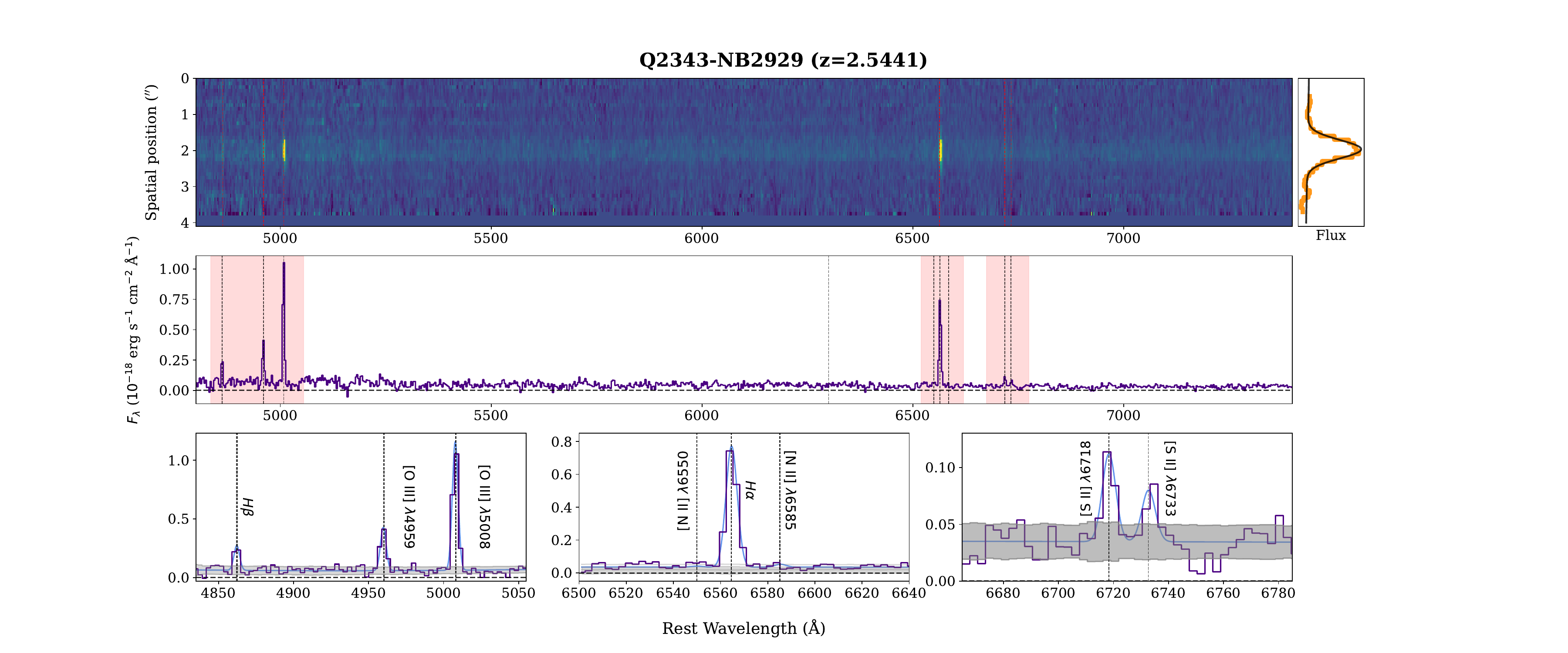}
   \hskip -2ex
   \caption{The rest-frame NIRSpec spectrum of Q2343-NB2929. \textit{Top:} 2D spectrogram of the source, with key emission lines denoted by red vertical lines. The panel on its right shows the 1D spatial distribution of the emission lines (orange line) and the gaussian fit used for optimal extraction. \textit{Middle:} Full spectral range of the G235M spectrum for the source. Dashed lines and pink shaded bands indicate the positions of emission lines highlighted in the bottom panels. The 1D extracted spectrum is shown in purple. \textit{Bottom:} The zoom-in panels of key spectral features. The Gaussian+linear continuum fit to the emission lines is shown in blue, and the error is shown as a gray region.}
   \label{fig:1D spectrum}
\end{figure*}

First, we identify the position of the galaxy in the 2D spectrogram (top panel in Figure \ref{fig:1D spectrum}) by extracting the portions of the 2D spectra near the brightest emission lines of the galaxy. These lines are determined after visual inspection of the 2D spectrograms and include H$\beta$, [O III] $\lambda$4959, [O III] $\lambda$5007, H$\alpha$, [S II] $\lambda$6716, and [S II] $\lambda$6731. For the CECILIA fLGBs, we also include He~I~$\lambda7065$, as it yielded significant detection. The wavelength extraction window for each emission line is 16 \AA\ in the rest-frame, which corresponds to approximately 1000 km/s, about three times the full width at half maximum (FWHM) of the typical line width in the dataset. We then estimate the 1D spatial distribution of these lines by calculating the mean of each pixel row. Subsequently, a Gaussian profile is fit to the 1D spatial profile, from which the center and sigma values are determined. The typical sigma is found to be $0\farcs16$.

Using the 1D spatial distribution profile of the emission lines, we perform optimal extraction of the 1D spectra, following the approach outlined by \citet{Horne1986}.


\subsubsection{Serendipitous Sources}
\label{sec:serendipitous_sources}


The process of extracting the serendipitous sources involves creating a 1D spatial profile for the emission lines of both the main and the serendipitous sources, based on their brightest emission lines, similarly as for the targeted sources. After that, Gaussian profiles are simultaneously fit to the spatial distributions of both the primary and serendipitous sources and their center and sigma values are calculated. Then, we proceed to fit these two Gaussian profiles, with their center and sigma values fixed to those extracted, to each column of data of the 2D spectrogram, for every pixel position in the wavelength direction — i.e., only the amplitudes of the Gaussians are allowed to vary with wavelength. Finally, from each column, a flux value is measured, and in doing so, we construct the respective 1D spectra of the serendipitous sources.

Because C31b and its primary source C31 are nearly spatially coincident in the 2D spectrogram, we also tested our method by performing a boxcar extraction of their combined spectrum and fitting a single model that accounts for the emission lines and redshifts of both sources. We found that the resulting emission line ratios agree with those obtained from the extraction method based on the gaussian spatial profile.




\subsection{Flux Corrections}
\label{sec:flux_corrections}

To address persistent flux calibration issues identified in the NIRSpec data reduction, we rescaled the G235M spectra using a relative flux calibration approach as described by \citetalias{stro2023} and \citetalias{2024ApJ...964L..12R}. Specifically, a low-order polynomial flux correction function $f_{\textrm{corr}}(\lambda)$ is estimated for each of the more luminous CECILIA sources in order to match the NIRSpec-measured continuum to the best-fit SED model for each source. This function is then multiplied by the measured spectrum to construct a flux-corrected spectrum.

\begin{figure}[t]
   \centering
   \includegraphics[width=0.47\textwidth]{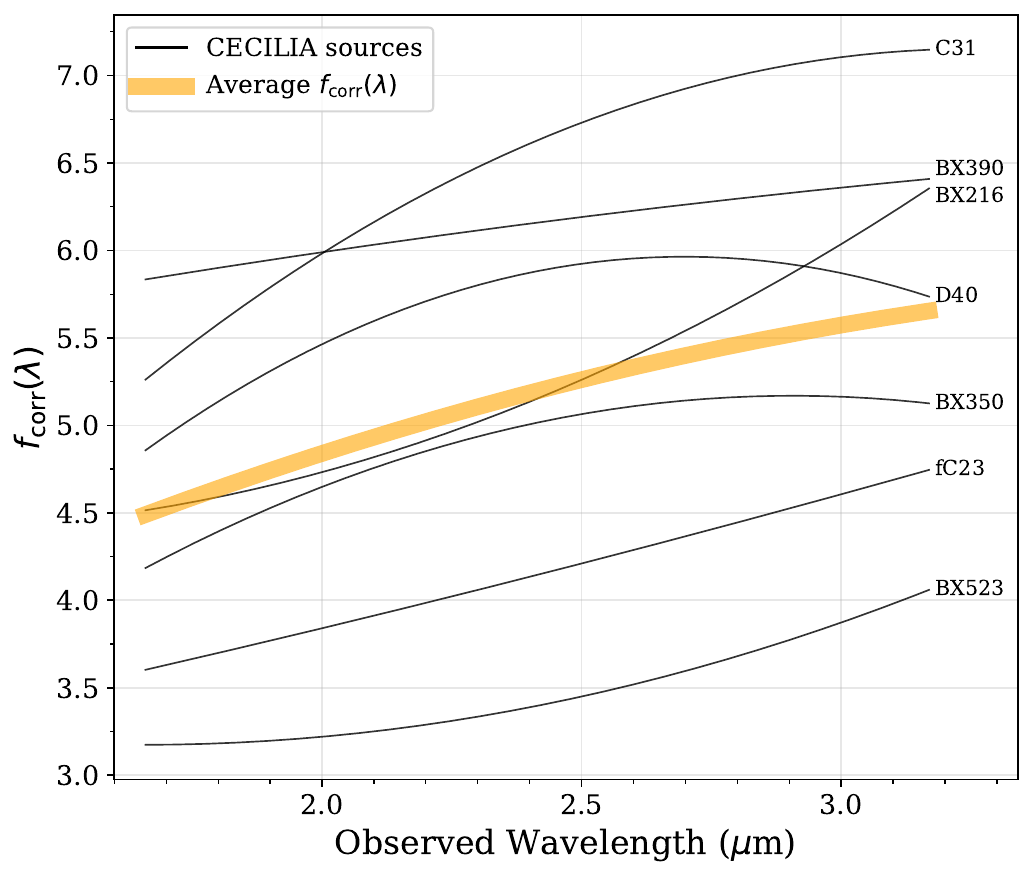}
   \caption{Flux correction functions $f_{\text{corr}}(\lambda)$ for galaxies in the overall CECILIA sample with well-measured SEDs, below-median luminosities, and below-median sizes. The black curves show the individual flux correction functions for these galaxies, while the thick orange curve shows the average flux correction function, constructed as the arithmetic mean of the individual $f_{\text{corr}}(\lambda)$ curves.}
   \label{fig:flux_corrections}
\end{figure}

The fLBG galaxies described in this work have individual flux-correction functions estimated as described by \citetalias{stro2023} and Rogers et al. (in preparation). However, the LAEs in our sample are typically undetected in continuum imaging and have less-secure SED models. As such, we construct an average flux correction function by the arithmetic mean of the flux correction functions for the subset of CECILIA galaxies with well-measured SEDs, below-median luminosities, and below-median sizes (Figure \ref{fig:flux_corrections}). This average flux correction function is applied to both the LAEs and the serendipitous sources. For the latter, we find that using the flux correction functions derived from their associated primary sources --- with which they share a slit --- yields $E(B-V)$ values and star formation rates that remain within the $1\sigma$ uncertainties of the reported measurements. The flux correction functions vary in absolute scaling by a factor of $\sim$2 at the wavelength of H$\alpha$, and the variations in shape produce a relative flux scaling uncertainty of $\sim$10\% in the H$\alpha$/H$\beta$ ratio. We also note that the absolute flux calibration for the serendipitous sources is uncertain since their centering in the slit is unknown. These systematic uncertainties in the flux calibration are not included in the uncertainties quoted for the individual line flux measurements but are included in the derived quantities such as star-formation rates and $E(B\!-\!V)$ values presented in this paper.

\subsection{Line Fitting}
\label{sec:line_fitting}

After extracting the 1D spectrum for each source, we proceed to fit a multi-component spectral model that includes Gaussian emission lines and a polynomial continuum. The final reduced rest-frame spectrum of the example source Q2343-NB2929 is plotted in Figure \ref{fig:1D spectrum}.

For fitting the emission lines, we use a Gaussian model for each line, and the line widths are tied together with a common velocity dispersion, $\sigma_v$. We set this parameter to an initial value of 70 km/s, with a minimum allowed value of 30 km/s. The line widths are tied through the expression for the width, $\sigma_\lambda = \frac{\sigma_v}{c} \times \lambda_{\textrm{cen}}$, where $\sigma_\lambda$ is in units of angstroms. 

Based on the residuals from this initial fit, we scale the flux uncertainties (and thus the corresponding weights) using the interquartile range (IQR) method \citep{Ilyushin2024} described below. These scaled uncertainties are then adopted for the subsequent, more precise fits performed within smaller spectral windows. In the windowed re-fits, the line width is allowed to vary independently in each spectral window to account for uncertainties in the instrumental resolution as a function of wavelength — all the fitted sections are less than 1000 \AA\ in width and include at least one fairly bright line that dominates the measured width within that region. Also, note that most of the lines measured for our objects are consistent with or narrower than the nominal instrumental resolution ($\sigma_v \sim 100$–$200$ km/s), indicating that our objects are physically smaller than the $0\farcs2$ slit widths of NIRSpec/MSA. Due to uncertainties in the effective resolution of the spectra, we do not attempt to correct for the instrumental resolution. We 
defer a more detailed analysis of line widths to future work.

Additionally, certain line flux ratios are fixed in our fits based on known physical relations. For instance, the flux ratio of [OIII] $\lambda4959$ and [OIII] $\lambda5007$ is fixed at a 1:3 ratio, as expected from atomic physics \citep{Osterbrock2006}. Similarly, the ratio between [NII] $\lambda6549$ and [NII] $\lambda6583$ is fixed at 1:3.
This ensures that the fits respect physical constraints on the emission line ratios. Moreover, the fitting process includes a quadratic component to account for the continuum in each spectral window.

The associated line flux uncertainties are calculated using Monte Carlo realizations. To estimate the flux uncertainties, we first extract the 1D flux error array produced by the JWST pipeline. This is done after confirming that our apertures, derived from the 1D spatial distribution profiles, are consistent with those used in its calculation. For the serendipitous sources, the 1D error arrays are derived from the standard deviation of the measured fluxes obtained through the initial weighted fitting of each column in the 2D spectrograms. To ensure the pipeline-produced error arrays reflect the typical spread in the data, we employ the IQR method for scaling the residuals between the 1D extracted flux and its fit. Outliers are filtered out based on calculated bounds derived from the IQR. Specifically, we determine the lower and upper bounds by subtracting 1.8 times the IQR from the first quartile (Q1) to establish the lower bound, and by adding 1.8 times the IQR to the third quartile (Q3) to establish the upper bound. Data points falling below the lower bound or above the upper bound are considered outliers and are excluded from the calculation of the error scaling factor, but they remain part of the spectral analysis. The resulting scaling factors, computed as the standard deviation of the filtered residuals relative to the original flux errors, have an average value of $\sim$1.5 across our sample.

With the scaled flux errors, we proceed to using Monte Carlo realizations to quantify the uncertainties in the emission-line flux measurements by generating multiple synthetic spectra and adding random Gaussian noise, scaled by the updated per-pixel flux errors, to the 1D spectrum. For each synthetic spectrum, we fit the spectrum and record the parameter values (flux, width, and center position) for the emission lines. Repeating this process 400 times, we create a distribution of parameter values for each emission line. The standard deviation of this distribution provides the estimate of the flux uncertainty for each emission line. 

Due to the faint nature of the sources and the shorter integration times of the G395M spectra, individual G395M spectra do not yield significant detections of emission lines for the majority of our individual continuum-faint sources, so we do not report measurements for individual sources in this band.

\subsection{Stack Spectra}
\label{sec:stack_spectra}

\begin{figure*}[h]
   \centering
   \includegraphics[width=0.9\textwidth]{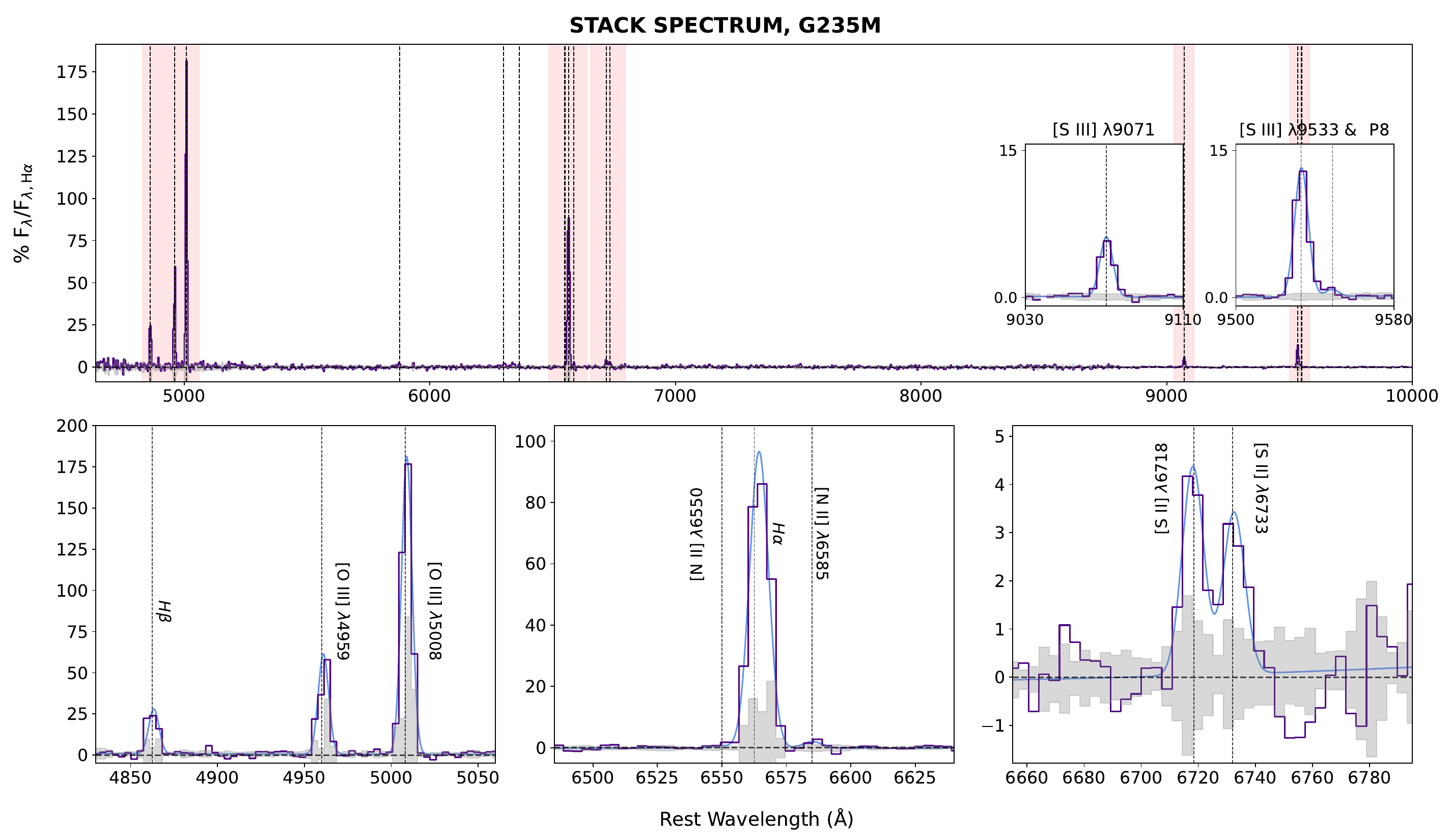}
   \caption{The rest-frame NIRSpec G235M stack spectrum (NB, fLBGs, and serendipitous sources). The purple line represents the 1D extracted spectrum, while the blue line illustrates the Gaussian+linear continuum fit to the emission lines. The gray shaded area indicates the error margin. A horizontal dashed black line at $y = 0$ marks the zero level. Black dashed vertical lines indicate the positions of key spectral features — most of which are shown in the bottom panels — as well as emission lines with $>2\sigma$ detections. Individual spectra are normalized by their integrated H$\alpha$ flux prior to stacking. The resulting stacked spectrum is then rescaled such that the H$\alpha$ peak of the fit model has a value of 100, allowing all other emission line peaks to be expressed as percentages of the H$\alpha$ peak.}
   \label{fig:stack_g235}
\end{figure*}

\begin{figure*}[h!]
   \centering
   \includegraphics[width=0.9\textwidth]{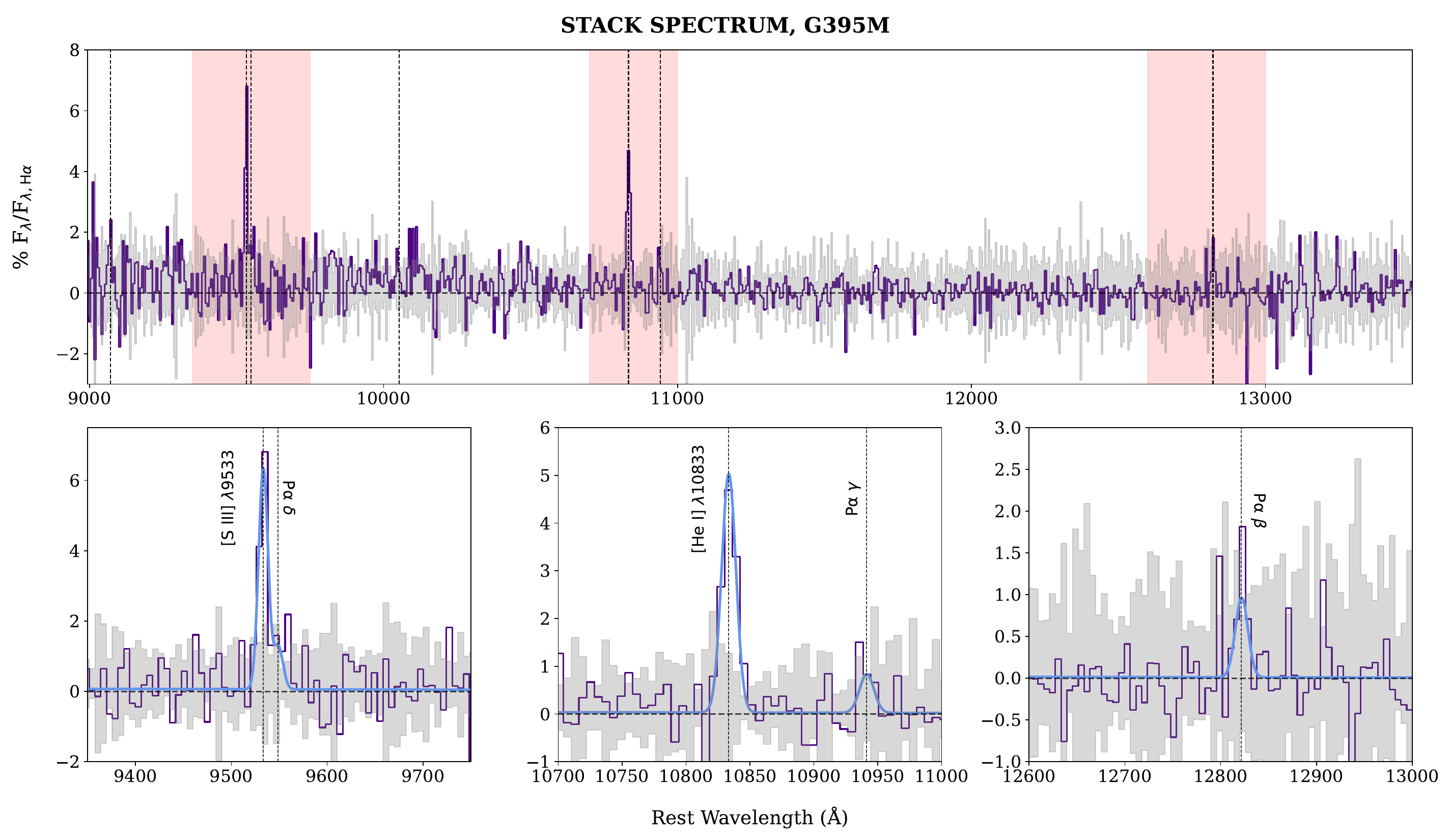}
   \hskip -2ex
   \caption{The rest-frame NIRSpec stack spectrum for the G395M segment. The spectrum is presented using the same scaling as in Figure~\ref{fig:stack_g235}.}
   \label{fig:STACKG395}
\end{figure*}

Stacking spectra enhances the signal-to-noise ratio, enabling the detection of faint emission lines and more accurate flux measurements. This approach reveals the average physical conditions of these galaxies, offering insights into their ionization and physical conditions that are unconstrained from individual spectra. To create the G235M stack spectrum (Figure \ref{fig:stack_g235}), the 1D spectra are interpolated onto a common wavelength grid, with a resolution equal to the typical rest-frame resolution of the individual spectra. Prior to interpolation, emission lines from the primary sources within the serendipitous spectra are masked to prevent contamination in the stacking process. Each spectrum is scaled by its H$\alpha$ flux and undergoes continuum subtraction using the polynomial components of the fitted model discussed in the previous section. We use sigma clipping to exclude 3$\sigma$ outliers at every rest wavelength. We compute the stack spectrum by taking the mean of the sigma-clipped fluxes. To construct the corresponding error spectrum, we generate 1000 bootstrap-resampled stacks and compute the standard deviation of flux values at each pixel. For emission line flux uncertainties, within each resampling loop, we fit the model to the resampled stack and save the resulting fluxes. This process provides 1000 flux measurements for each emission line, from which we calculate the standard deviation to determine the flux uncertainty for each line. The same approach is used for the G395M stack spectrum (Figure \ref{fig:STACKG395}).

A comprehensive list of emission lines measured from the stack G235M and G395M spectra is displayed along with their vacuum wavelengths in Table \ref{tab:G235_fluxes}, with flux measurements relative to H$\alpha$. For the G395M stack, we fit seven emission lines, of which, [S~III]$\lambda$9531 and He~I$\lambda$10830 exhibit a marginally significant detection exceeding \(2\sigma\).

\begin{deluxetable}{lcc}
\tablecaption{Flux measurements relative to H$\alpha$ for the CECILIA Faint Stack spectrum. \label{tab:G235_fluxes}} 
\tablehead{
    \colhead{} & 
    \colhead{G235M} &
    \colhead{} \\ [-0.2cm]
    \colhead{Emission Line} &
    \colhead{Vacuum Wavelength (\AA)} &
    \colhead{Flux Relative to H$\alpha$} 
}
\startdata
H$\beta$ $\lambda$4861 & 4862.69 & 0.225 $\pm$ 0.064 \\
\textnormal{[O III]} $\lambda$4959 & 4960.30 & 0.512 $\pm$ 0.140 \\
\textnormal{[O III]} $\lambda$5007 & 5008.24 & 1.538 $\pm$ 0.419 \\
He I $\lambda$5876 & 5877.25 & 0.023 $\pm$ 0.010 \\
\textnormal{[O I]} $\lambda$6300 & 6302.89 & 0.023 $\pm$ 0.006 \\
\textnormal{[O I]} $\lambda$6364 & 6363.78 & 0.015 $\pm$ 0.006 \\
\textnormal{Si II} $\lambda$6373 & 6373.12 & <0.003 \\
\textnormal{[N II]} $\lambda$6550 & 6549.86 & 0.007 $\pm$ 0.003 \\
H$\alpha$ $\lambda$6563 & 6564.58 & 1.000 $\pm$ 0.183 \\
\textnormal{[N II]} $\lambda$6583 & 6585.27 & 0.022 $\pm$ 0.009 \\
\textnormal{[S II]} $\lambda$6716 & 6718.29 & 0.046 $\pm$ 0.012 \\
\textnormal{[S II]} $\lambda$6731 & 6732.67 & 0.036 $\pm$ 0.010 \\
He I $\lambda$7065 & 7067.14 & <0.079\\
\textnormal{[Ar III]} $\lambda$7136 & 7137.77 & <0.087\\
\textnormal{[O II]} $\lambda$7320 & 7321.94 & <0.078\\
\textnormal{[O II]} $\lambda$7331 & 7332.21 & <0.069\\
\textnormal{[Ar III]} $\lambda$7751 & 7753.19 & <0.065\\
\textnormal{[S III]} $\lambda$9069 & 9068.60 & 0.056 $\pm$ 0.008 \\
\textnormal{[S III]} $\lambda$9531 & 9530.60 & 0.127 $\pm$ 0.017 \\
P8 $\lambda$9546 & 9548.80 & 0.007 $\pm$ 0.003\\
\hline \\ 
\multicolumn{3}{c}{G395M} \\ 
Emission Line & Vacuum Wavelength (\AA) & Flux \\ 
\hline 
\siii\ $\lambda$9069 & 9070.52 & <0.068 \\ 
\siii\ $\lambda$9531 & 9533.10 & 0.060 $\pm$ 0.021 \\ 
P8 $\lambda$9546 & 9548.98 & <0.075 \\ 
P7 $\lambda$10049 & 10053.15 & <0.029  \\ 
\ion{He}{1} $\lambda$10830 & 10832.55 & 0.053 $\pm$ 0.022  \\ 
P6 $\lambda$10938 & 10940.50 & <0.068  \\ 
P5 $\lambda$12818 & 12821.80 & <0.071  \\ 
\enddata

\end{deluxetable}

\setlength{\tabcolsep}{3.5pt}
\begin{deluxetable*}{lccccccccc}
\tablecaption{Flux measurements for the G235M spectra.} 
\tablehead{
    \colhead{Emission Line} &
    \colhead{NB2929} &
    \colhead{NB2089 } &
    \colhead{NB2875} &
    \colhead{NB2571}&
    \colhead{fBM47} &
    \colhead{fBM40}&
    \colhead{fC23} &
    \colhead{C31b}&
    \colhead{BX587b}
}
\startdata
H$\beta$ $\lambda$4861 & $1.00 \pm 0.16$ & $1.91 \pm 0.09$ & $2.30 \pm 0.10$ & $1.56 \pm 0.14$ & \dots  & \dots  & \dots  & $1.56 \pm 0.12$ & $0.64 \pm 0.11$ \\
\textnormal{[O III]} $\lambda$4959 & $1.77 \pm 0.05$ & $1.54 \pm 0.03$ & $2.14 \pm 0.03$ & $2.89 \pm 0.05$ & \dots  & \dots  & \dots  & $4.11 \pm 0.30$ & $1.45 \pm 0.05$ \\
\textnormal{[O III]} $\lambda$5007 & $5.32 \pm 0.15$ & $4.62 \pm 0.07$ & $6.43 \pm 0.10$ & $8.68 \pm 0.14$ & \dots  & \dots  & \dots  & $12.32 \pm 0.90$ & $4.34 \pm 0.14$ \\
He I $\lambda$5876 & $<$0.17 & $0.30 \pm 0.04$ & $0.19 \pm 0.04$ & $<$0.29 & $0.79 \pm 0.08$ & $0.99 \pm 0.06$ & \dots  & $0.29 \pm 0.07$ & $0.35 \pm 0.08$ \\
\textnormal{[O I]} $\lambda$6300 & $<$0.15 & $<$0.13  & $<$0.18 & \dots  & $0.30 \pm 0.08$ & $0.56 \pm 0.05$ & \dots  & $<$0.19 & $<$0.25 \\
\textnormal{[O I]} $\lambda$6364 & $<$0.23 & $<$0.19 & $<$0.07 & \dots  & $<$0.21 & $0.19 \pm 0.06$ & \dots  & $<$0.11 & $<$0.10 \\
\textnormal{Si II} $\lambda$6373 & $<$0.14 & $<$0.08 & $<$0.17 & \dots  & $<$0.19 & $0.19 \pm 0.05$ & \dots  & $<$0.18 & $<$0.11\\
\textnormal{[N II]} $\lambda$6549 & $<$0.09 & $<$0.02 & $<$0.03 & $<$0.08 & $0.22 \pm 0.03$ & $0.63 \pm 0.02$ & $0.70 \pm 0.02$ & $0.07 \pm 0.02$ & $<$0.05\\
H$\alpha$ $\lambda$6563 & $4.43 \pm 0.10$ & $7.19 \pm 0.05$ & $6.74 \pm 0.07$ & $4.63 \pm 0.08$ & $22.45 \pm 0.14$ & $28.96 \pm 0.13$ & $20.94 \pm 0.10$ & $5.54 \pm 1.05$ & $4.83 \pm 0.12$ \\
\textnormal{[N II]} $\lambda$6583 & $<$0.28 & $<$0.06 & $<$0.12 & $<$0.27 & $0.67 \pm 0.09$ & $1.89 \pm 0.06$ & $2.34 \pm 0.06$ & $0.22 \pm 0.07$ & $<$0.23 \\
\textnormal{[S II]} $\lambda$6716 & $0.48 \pm 0.09$ & $<$0.16 & $<$0.06 & $<$0.17 & $1.39 \pm 0.09$ & $2.36 \pm 0.07$ & $2.06 \pm 0.06$ & $0.24 \pm 0.06$ & \dots \\
\textnormal{[S II]} $\lambda$6731 & $0.28 \pm 0.08$ & $<$0.12 & $<$0.07 & $<$0.18 & $0.98 \pm 0.09$ & $1.87 \pm 0.05$ & $1.52 \pm 0.06$ & $0.17 \pm 0.06$ & \dots \\
He I $\lambda$7065 & $<$0.08 & $0.11 \pm 0.04$ & $<$0.14 & $0.07 \pm 0.03$ & $0.40 \pm 0.08$ & $0.23 \pm 0.05$ & $0.14 \pm 0.07$ & $<$0.16 & $0.29 \pm 0.09$ \\
\textnormal{[Ar III]} $\lambda$7136 & $<$0.14 & $0.09 \pm 0.04$ & $<$0.06 & $<$0.20 & $0.66 \pm 0.08$ & $0.67 \pm 0.05$ & $0.50 \pm 0.06$ & $0.28 \pm 0.10$ & $0.25 \pm 0.08$ \\
\textnormal{[O II]} $\lambda$7320 & $<$0.30 & $<$0.15 & $<$0.10 & $<$0.05 & $<$0.23 & $0.32 \pm 0.04$ & $0.28 \pm 0.05$ & $<$0.24 & $<$0.20 \\
\textnormal{[O II]} $\lambda$7331 & $<$0.19 & $<$0.07 & $<$0.15 & $<$0.25 & $<$0.05 & $0.30 \pm 0.04$ & $0.20 \pm 0.05$ & $<$0.08 & $<$0.13 \\
\textnormal{[Ar III]} $\lambda$7751 & $<$0.17 & $<$0.17 & $<$0.05 & $0.20 \pm 0.10$ & $<$0.26 & $0.17 \pm 0.04$ & $<$0.21 & $<$0.14 & $<$0.32 \\
\textnormal{[S III]} $\lambda$9069 & \dots & \dots & \dots & \dots & \dots & $1.41 \pm 0.05$ & $1.38 \pm 0.06$ & \dots & \dots \\
\textnormal{[S III]} $\lambda$9531 & \dots & \dots & \dots & \dots & $2.19 \pm 0.08$ & $3.63 \pm 0.06$ & $3.21 \pm 0.06$ & \dots & \dots \\
P8 $\lambda$9546 & \dots & \dots & \dots & \dots & $<$0.20 & $0.31 \pm 0.04$ & $0.14 \pm 0.05$ & \dots & \dots \\
\enddata 
\label{tab:allG235_fluxes}
\tablecomments{Line fluxes in $10^{-18}~\mathrm{erg~s^{-1}~cm^{-2}}$ measured from the NIRSpec rest-optical data. These fluxes are not dust-corrected. The uncertainties are statistical and do not take into account uncertainties in the flux correction functions described in Section \ref{sec:flux_corrections}.}
\end{deluxetable*}


\section{Results}

\subsection{Dust Attenuation}
\label{section_dust_attenuation}
To account for dust attenuation in our galaxy sample, we calculate the Balmer Decrement (BD; H$\alpha$/H$\beta$) from the observed fluxes of the H$\alpha$ and H$\beta$ emission lines. We derive nebular $E(B-V)$ values for each galaxy using an intrinsic H$\alpha$/H$\beta$ ratio of 2.89 and the \citet{1989ApJ...345..245C} Galactic attenuation law with $R_V = 3.1$. Other attenuation laws calibrated on galaxies at Cosmic Noon yield similar results \citep[e.g.,][]{redd2020}. The individual $E(B-V)$ values are reported in Table~\ref{t:galaxies}. For the CECILIA fLBGs, $E(B-V)$ values are calculated from previous Keck/MOSFIRE measurements \citepalias{stro2017}, due to the lack of NIRSpec H$\beta$ coverage.

Our sample spans a wide range of dust attenuation, with $E(B-V)$ values from 0.05 to 0.95 (Figure~\ref{fig:ebv_properties}) and a median of 0.23. We measure $E(B-V) = 0.43 \pm 0.26$ from the stacked spectrum.

These extinction values are broadly consistent with previous studies of $z \sim 2$ LAEs. For instance, \citet{Guaita2011}, \citet{Nakajima2012}, and \citet{Oteo2015} report typical values of $E(B-V)_\star \sim 0.2$--$0.3$ based on SED fitting. Using the same Balmer-decrement method employed here, \citetalias{Trainor2016} found $E(B-V) = 0.06 \pm 0.12$ for 60 faint LAEs from the KBSS-Ly$\alpha$ survey at $z \approx 2.56$, while \citet{Sanders2024} reported values ranging from 0.06 to 0.34 for their $z \sim 2$--3 sample. Both are consistent with the extinction values measured for our faintest LAEs.

\subsection{Star formation rates}
\label{star_formation_rates_section}
We measure the SFRs for all the narrow-band and serendipitous sources using the dust-corrected H$\alpha$ luminosities. We estimate the extinction as a multiplicative factor at the rest-wavelengths of H$\alpha$, which is assumed to be the vacuum wavelength. To calculate the SFRs for all the sources in Table \ref{t:galaxies}, we use the formula:
\begin{equation}
\text{SFR} = 10^{-41.68} \, \frac{L_{\mathrm{H}\alpha}}{\mathrm{erg} \, \mathrm{s}^{-1}} \, \left[\frac{\mathrm{M}_{\odot}}{\mathrm{yr}}\right],
\label{eq:sfr_ha}
\end{equation}
where $L_{\mathrm{H}\alpha}$ is the H$\alpha$ luminosity. Our SFR calibration follows a form similar to the widely used relation from \citet{Kennicutt1998}, but with a normalization specifically derived from metal-poor BPASS models at an absolute stellar metallicity of $Z_* = 0.001$ (7\% Solar), as described in \citet{korh2025}. This conversion factor is appropriate for the nebular oxygen abundances ($\sim$20--30\%~Solar; Sections~\ref{Photoionization}--\ref{sec:comparison_to_direct}) expected for our sample under the assumption of $\alpha$-enhanced abundance patterns, with O/Fe ratios of $\sim$2--4~$(\mathrm{O/Fe})_\odot$ previously measured in star-forming galaxies at $z \sim 2$--3 \citep[e.g.,][]{stei2016, stro2018, cull2019, topp2020}. This low-metallicity calibration is well suited to our faint, metal-poor sample and similar LAE populations (e.g., the KBSS-Ly$\alpha$ sample), but may not accurately reflect the SFR-to-$L_{\mathrm{H}\alpha}$ ratio of more metal-rich or heterogeneous galaxy samples such as the full KBSS and MOSDEF surveys, which span a wider range of metallicities and ionizing conditions.

To estimate the uncertainties on the SFRs, we implement a Monte Carlo approach. For each spectrum, 400 realizations of the H$\alpha$ and H$\beta$ fluxes are drawn from normal distributions consistent with their respective measurement uncertainties. For each pair, we compute $E(B-V)$ (as described in Section~\ref{section_dust_attenuation}) and use it to apply a dust correction to the corresponding H$\alpha$ flux. This results in 400 dust-corrected H$\alpha$ luminosities and hence 400 SFRs per galaxy. We adopt the 16\textsuperscript{th} and 84\textsuperscript{th} percentiles of the resulting distribution as the lower and upper bounds on the SFR uncertainties.

\begin{figure}[h]
   \centering
   \includegraphics[width=0.47\textwidth]{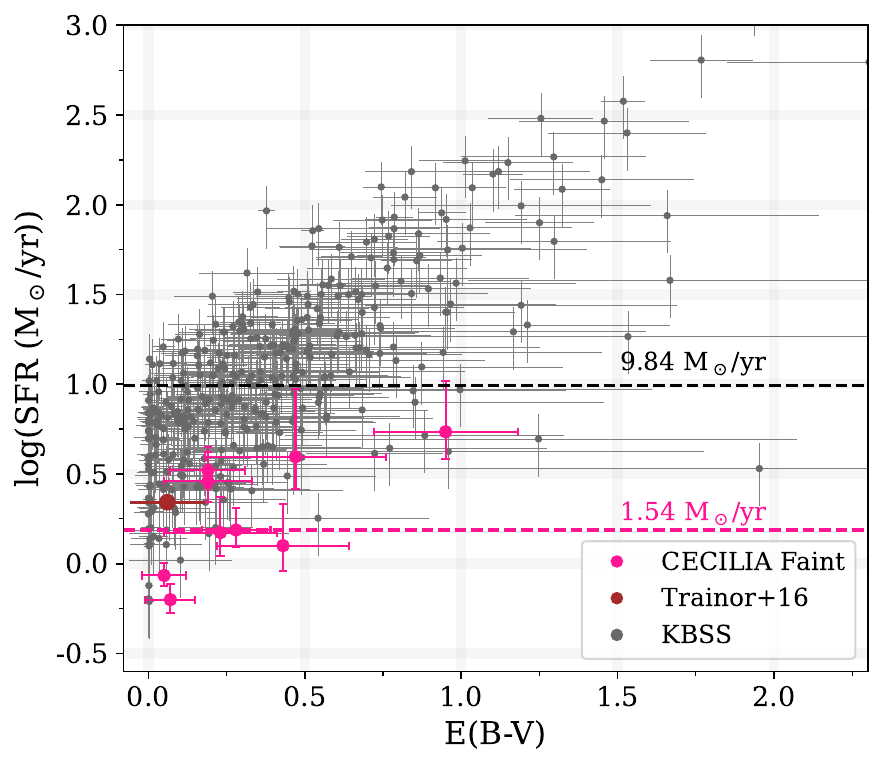}
   \caption{Log(SFR) versus $E(B{-}V)$ for the CECILIA-faint sample (pink circles), the \citetalias{Trainor2016} KBSS-Ly$\alpha$ stack point (brown circle), and KBSS LBGs from \citetalias{stro2017} (gray circles). The horizontal dashed pink line indicates the median SFR of the CECILIA-faint sample, while the dashed black line marks the median SFR for the KBSS points. A positive correlation is seen between dust attenuation and SFR across the samples.}
   \label{fig:ebv_properties}
\end{figure}

The SFRs in our continuum-faint sample range from 0.63 to 5.43~M$_{\odot}$~yr$^{-1}$, with a median of 1.54~M$_{\odot}$~yr$^{-1}$. As shown in Figure~\ref{fig:ebv_properties}, we observe an apparent positive trend between $E(B-V)$ and SFR, such that galaxies with higher SFRs also tend to exhibit greater dust attenuation. However, we caution that this correlation may partly reflect an observational bias, since both quantities are derived from the same underlying H$\alpha$ and H$\beta$ measurements.


Compared to other $z\sim2-3$ galaxy samples, our continuum-faint galaxies show systematically lower SFRs, reflecting both their intrinsically low luminosities and our targeting strategy. For instance, \citetalias{Trainor2016} reported a median SFR of 5.3~M$_\odot$~yr$^{-1}$ for 28 LAEs using the \citet{Kennicutt1998} calibration, while the MOSDEF survey \citep{Shapley2019} and the KBSS LBG sample \citepalias{stro2017} found significantly higher values. The KBSS LBGs have a median SFR of 24~M$_\odot$~yr$^{-1}$. Both studies adopt the \citet{Kennicutt1998} relation with a \citet{Chabrier2003} initial mass function (IMF). For consistency, we recalculate SFRs for these comparison samples using our low-metallicity calibration from Equation~\ref{eq:sfr_ha}. This yields revised median SFRs of 2.2~M$_\odot$~yr$^{-1}$ for the KBSS-Ly$\alpha$ LAEs — slightly higher than our own median — and 10.0~M$_\odot$~yr$^{-1}$ for the KBSS LBGs, which remains significantly elevated relative to our sample. We further note that if we used a metallicity-dependent conversion factor for the KBSS LBGs as in \cite{korh2025}, we would find an even higher star-formation rate for them, owing to the higher average metallicity of that population.

These comparisons are illustrated in Figure~\ref{fig:ebv_properties}, where SFRs for all datasets are derived using the same conversion from Equation~\ref{eq:sfr_ha}. 
These star formation rates, along with the faint continuum luminosities and unresolved velocity dispersions, are consistent with stellar masses that are a factor of 10 to 100 smaller than those typical of KBSS and MOSDEF galaxies. A more detailed analysis of the SFR–stellar mass relation for these galaxies will be presented in future work.

\subsection{Electron Density}
\label{sec:electron_densities}

We estimate electron densities using the [S~II]~$\lambda6717/\lambda6731$ ratio at a fixed electron temperature of $T_e = 10^4$~K, following the methodology of \citetalias{2024ApJ...964L..12R}. Only fBM40 and fC23 have [S~II] $\lambda6717/\lambda6731$ detected sufficiently well to constrain their ratio at high signal-to-noise, with their inferred densities being $n_e = 205^{+39}_{-55}$~cm$^{-3}$ and $n_e = 95^{+93}_{-48}$~cm$^{-3}$, respectively.

The resulting electron densities are slightly below the typical densities found in $z \sim 2$ galaxies ($\sim250$~cm$^{-3}$; \citealt{sand2016}; \citetalias{stro2017}), and fall within the low-density regime where abundance measurements based on collisionally excited lines are largely insensitive to the adopted $n_e$. Both the stack and the individual sources are consistent with $n_e \lesssim 200$~cm$^{-3}$. More precise measurements would be necessary to establish whether intrinsically faint galaxies at $z > 2$ have systematically lower electron densities than their more luminous counterparts.

\subsection{Diagnostic Emission-line  Ratio Diagrams}
\label{Diagnostic_section}
\begin{figure*}[h]
   \centering
   \includegraphics[width=0.85\textwidth]{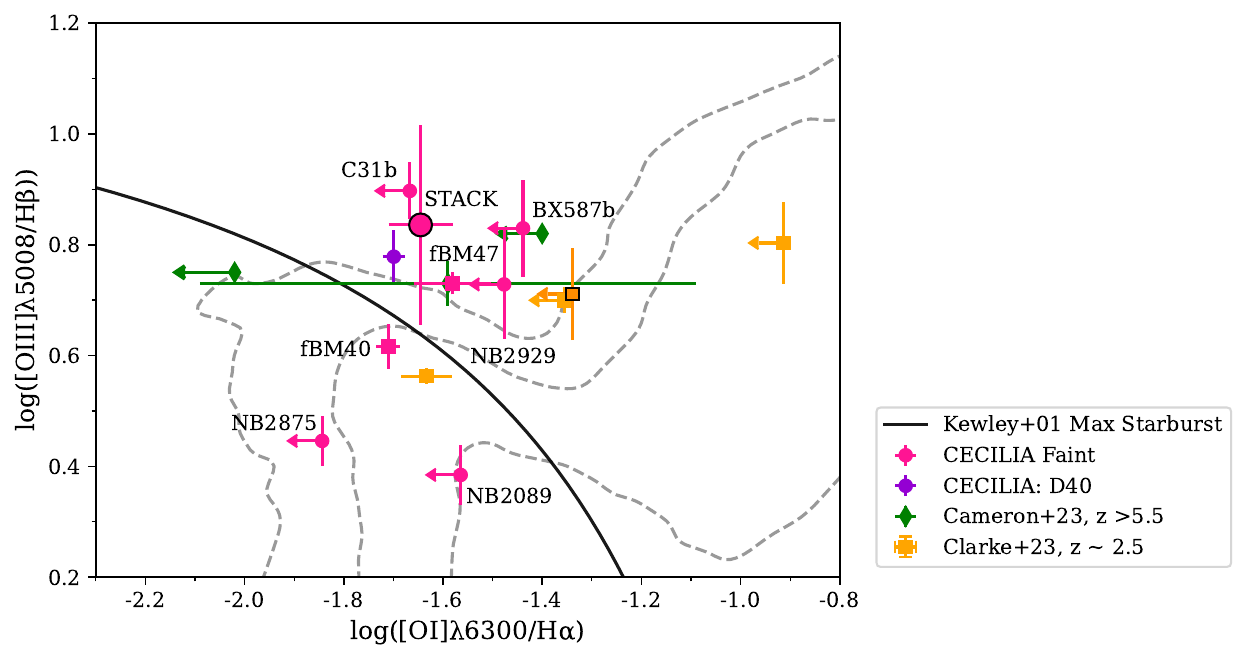}
   \caption{O1-BPT \citep{Veilleux1987} showing SDSS $z \sim 0$ galaxies (gray contour lines), CECILIA faint LAEs (pink circles with 2$\sigma$ limits), CECILIA faint LBGs (pink squares), CECILIA D40 (purple circle), Lyman Break galaxies from \cite{Cameron2023} (green diamonds), and SF galaxies at $z \sim 2.5$ (orange squares; stacked point with black outline) from \cite{Clarke2023}. The solid black line represents the theoretical starburst classification curve from \cite{Kewley2001}, separating SF galaxies from objects with other ionization mechanisms.}
   \label{fig:o1bpt}
\end{figure*}

As mentioned in Section \ref{Introduction}, emission-line diagnostic diagrams are commonly used to classify galaxies based on their ionization properties by comparing ratios of neighboring strong emission lines such as H$\beta$ and [O~III]$\lambda$5007. Several such ratios are defined in Table \ref{t:emission}. While the original diagram involving O3 and N2 was introduced by \citet{Baldwin1981}, additional diagnostics involving S2 and O1 were later developed by \citet{Veilleux1987}. These diagrams differentiate between galaxies whose gas is ionized by star formation, AGN, and low ionization emission-line regions (LINERs; see the review by \citealt{kewley2019} for more information). The most commonly used separation curves for these classifications in the BPT diagrams were proposed by \citet{Kewley2001, Kewley2006} and \citet{Kauffmann2003}; see curves in Figures \ref{fig:n2bpt}, \ref{fig:s2bpt}, and \ref{fig:o1bpt}. Galaxies at $z \gtrsim 2$ are observed to exhibit systematic shifts in these diagnostic diagrams compared to their low-redshift counterparts, such as those from the Sloan Digital Sky Survey (SDSS; \citealt{Wang2018, Garg2022, Scholtz2023}). These evolutionary trends suggest fundamental differences in the physical conditions of high-redshift galaxies, which will be discussed in the subsequent sections.

\subsubsection{N2-BPT Diagram}
\label{sec:N2_section}
The N2-BPT diagram is particularly effective in distinguishing star-forming (SF) galaxies from those with AGN activity, separating into two clear tracks for low-redshift galaxies \citep[e.g.,][]{Kauffmann2003, trem2004, Kewley2006}. The placement of SF galaxies on the N2-BPT diagram is primarily governed by metallicity, with low-metallicity H II regions typically exhibiting high O3 and low N2 values in the upper left, while metal-rich systems occupy the lower right \citep[e.g.,][]{Andrews2013}. 

\begin{deluxetable}{lc}
\tablecaption{Emission Line Ratios \label{t:emission}}
\tablewidth{0pt}
\tablehead{ 
  \colhead{Ratio}	&
  \colhead{Definition}
}
\startdata
O3 	& log([O III] $\lambda$5007/H$\beta$) \\ 
N2 	& log([N II] $\lambda$6583/H$\alpha$) \\
S2 	& log([S II] $\lambda\lambda$6716, 6731/H$\alpha$) \\
O1 	& log([O I] $\lambda$6300/H$\alpha$) \\ 
O3N2 & O3 $-$ N2 \\
\enddata
\end{deluxetable}

\begin{figure*}[t]
   \centering
   \includegraphics[width=1\textwidth]{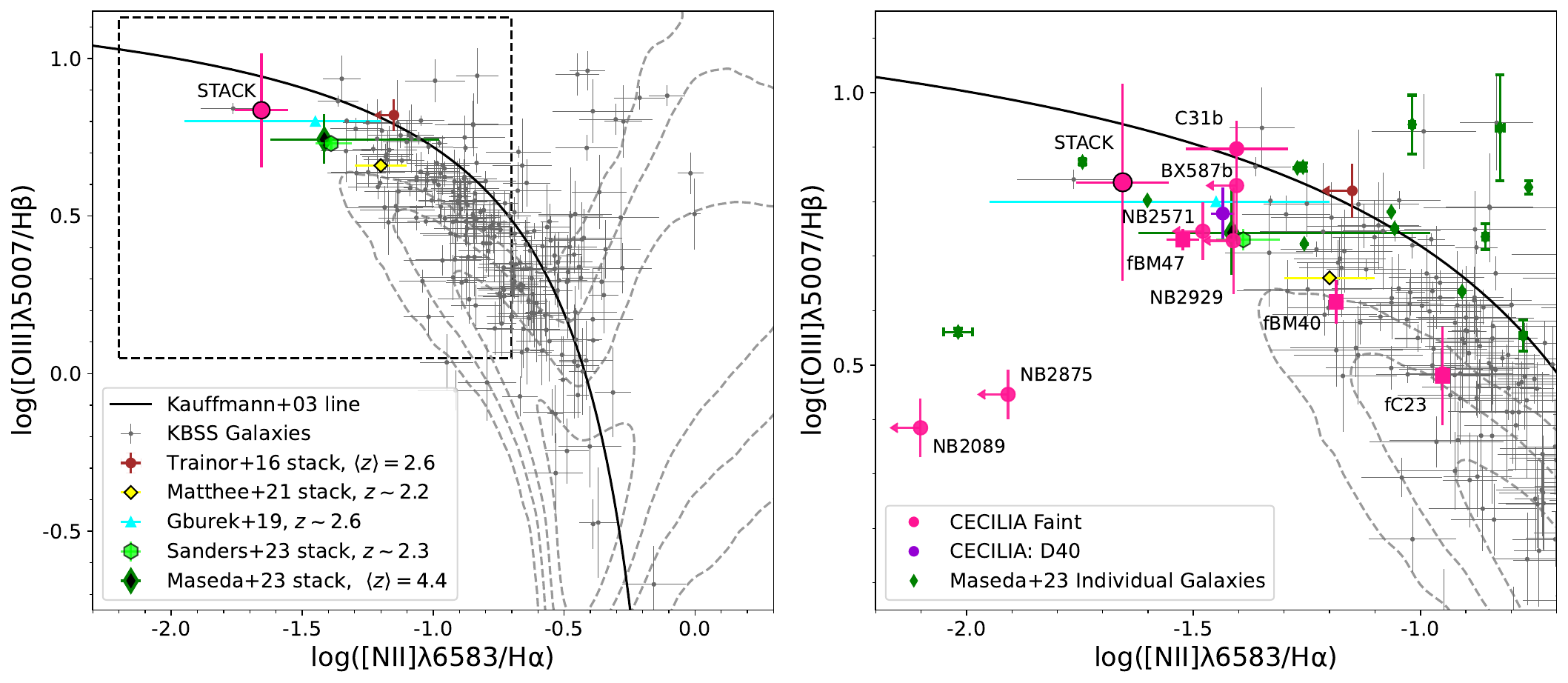}
   \caption{\textit{Left:} N2-BPT diagram \citep{Baldwin1981} with SDSS $z\sim0$ galaxies (gray, contour lines), KBSS LBGs (\citetalias{stro2017}; gray dots with 3 $\sigma$ detections for every line), a lensed dwarf galaxy \citep{Gburek2019}, stacks of faint LAEs (\citetalias{Trainor2016}; \citealt{Matthee2021, Maseda2023}), stack of high-$z$ galaxies \citep{Sanders2023}, and the CECILIA stack of the faint galaxies (pink point). The solid black line is the classification curve used by \cite{Kauffmann2003} as a lower limit for finding AGN. The dashed box highlights the region shown in the zoomed-in version. \textit{Right:} Zoomed-in N2-BPT diagram showing individual CECILIA faint galaxies (pink points; 2$\sigma$ limits, with squares indicating fLBGs), the $z\sim3$ SF galaxy Q2343-D40 observed with JWST/NIRSpec as part of the CECILIA program (purple point), and individual data points from \citet{Maseda2023}.}
   \label{fig:n2bpt}
\end{figure*}

At $z\sim2-3$, SF galaxies follow a similarly tight locus as their low-redshift counterparts but are systematically offset from the $z\sim0$ relation (e.g., \citealt{stei2014, Shapley2015, Laigle2016}; \citetalias{stro2017}). \citet{Law2021} confirmed that despite this offset, SF galaxies and AGNs at high redshift still occupy distinct sequences.

Figure~\ref{fig:n2bpt} shows the N2-BPT diagram for our continuum-faint galaxies, including four narrow-band selected LAEs, two serendipitous sources, and three continuum-faint CECILIA fLBGs. For most galaxies, the relevant emission lines are measured from JWST/NIRSpec spectra; the CECILIA fLBGs are instead observed with Keck/MOSFIRE \citepalias{stro2017}, as [O~III]$\lambda5007$ and H$\beta$ fall outside the NIRSpec coverage at their redshifts.

\begin{figure*}[b]
   \centering
   \includegraphics[width=0.8\textwidth]{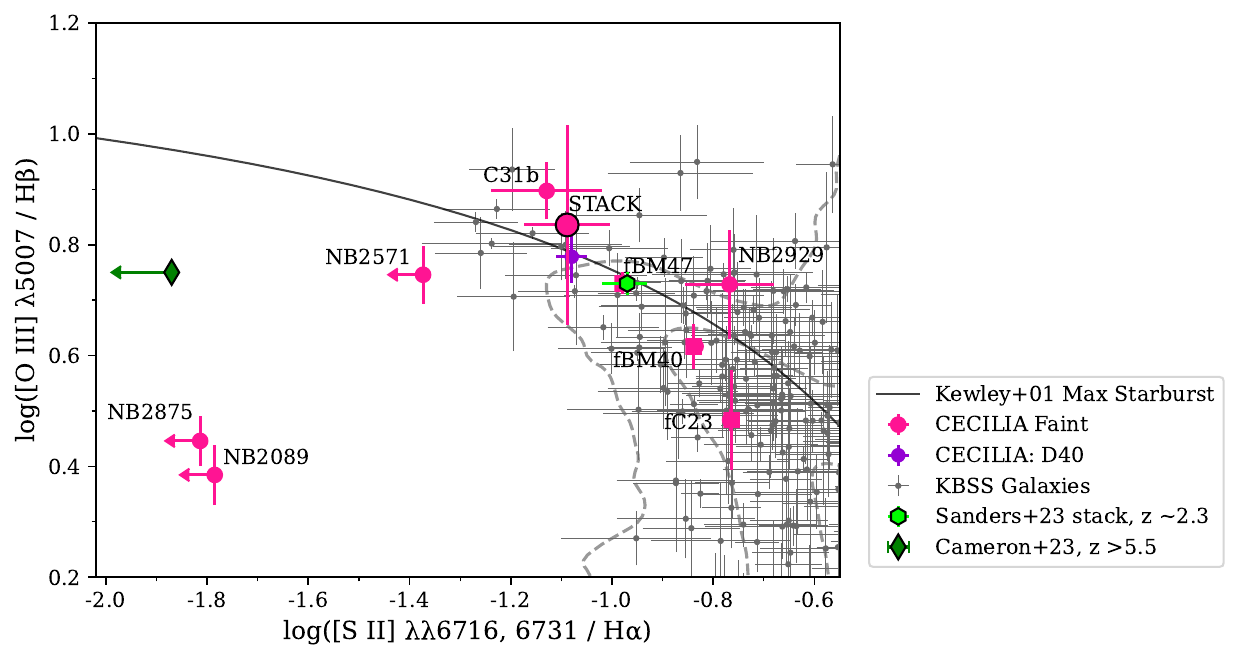}
   \caption{S2-BPT \citep{Veilleux1987} showing SDSS $z \sim 0$ galaxies (gray contour lines), KBSS LBGs from (\citetalias{stro2017}; gray dots with 3$\sigma$ detections for every line), a stack of high-$z$ galaxies from \cite{Sanders2023}, CECILIA faint LAEs (pink circles with 2$\sigma$ limits), CECILIA D40 (purple circle), a Lyman Break galaxy from \cite{Cameron2023}, and CECILIA faint LBGs (pink squares). The solid black line is the classification curve used by \cite{Kewley2001} to divide the theoretical starburst region from objects of other types of excitation.}
   \label{fig:s2bpt}
\end{figure*}

To place our results in context, we show comparison samples from the local SDSS population \citep[DR7;][]{Abazajian2009} and the KBSS LBG sample \citepalias{stro2017}, shown as gray points in Figures~\ref{fig:n2bpt}–\ref{fig:s2bpt}. The KBSS galaxies fall between the SDSS locus and the \citet{Kauffmann2003} AGN boundary, with their displacement attributed to harder stellar ionizing spectra rather than AGN contamination (see discussion in \citetalias{stro2017}). We also include Q2343-D40 (hereafter D40; \citetalias{2024ApJ...964L..12R}), a well-studied galaxy drawn from the KBSS and observed as part of the CECILIA program.

Most CECILIA continuum-faint galaxies exhibit low N2 and high O3 values, consistent with low-metallicity, high-ionization H~II regions. Of note, the fLBGs appear to follow the lower envelope of the KBSS LBG distribution, while the NB-selected sources occupy a somewhat distinct sequence in the diagram. Their positions on the N2-BPT diagram align with those of other faint high-redshift galaxies, including a lensed dwarf at $z\sim2.59$ \citep{Gburek2019} and stacked spectra of LAEs at $z\sim2-3$ (\citetalias{Trainor2016}; \citealt{Matthee2021, Sanders2023}), all near the upper left of the SDSS SF locus.

Two of the NB sources (NB2089, NB2875), which have the lowest N2 upper limits and among the faintest UV magnitudes in our sample, also exhibit lower O3 values compared to the other CECILIA Faint galaxies. NB2875, in particular, was identified by \citetalias{Trainor2016} as part of a population of low-O3, continuum-undetected LAEs in the KBSS-Ly$\alpha$ sample. 
\citetalias{Trainor2016} interpreted this behavior as evidence of very low metallicities using photoionization models, which show how O3 increases with decreasing metallicity until a turnover occurs $12 + \log(\mathrm{O/H}) \approx 8$. We explore this trend further using theoretical models in Section~\ref{Photoionization}.

\subsubsection{S2-BPT Diagram}
\label{sec:S2_section}

Figure \ref{fig:s2bpt} shows the S2-BPT diagram, along with the classification curve used by \citet{Kewley2001} to separate the theoretical starburst region from objects of other excitation types. As in the N2-BPT diagram, SF galaxies occupy a well-defined locus in S2 space. High-redshift galaxies from previous studies, including the $z \sim 2.3$ SF galaxy stack from \citet{Sanders2023} and the $z > 5.5$ Lyman break galaxy from \citet{Cameron2023}, also fall within the SF region of the S2-BPT diagram.

We find that all the CECILIA Faint sample galaxies are consistent with the SF region. Again, the S2-BPT shows that the $z \gtrsim 2$ galaxies generally track the SDSS locus, except for NB2089, NB2875, and the higher-$z$ Lyman break galaxy from \citet{Cameron2023}, which are offset toward lower O3 at fixed S2 compared to local SDSS galaxies. The relatively low S2 and O3 values of these three outliers again are consistent with the expected decline in O3 at $12 + \log(\text{O/H}) \lesssim 8.0$.

\subsubsection{O1-BPT Diagram}
Lastly, we present the first analysis of [O~I] emission in a sample of $L \ll L_\ast$ $z\sim2-3$ galaxies. Figure \ref{fig:o1bpt} shows the O1-BPT diagram with the CECILIA continuum-faint galaxies. Very few studies have reported [O~I] ratios at $z>2$ even for more luminous galaxies, so we also display comparison samples from $z\sim2.5$ \citep{Clarke2023}, $z>5.5$ \citep{Cameron2023}, and the general SDSS distribution.

The individual CECILIA galaxies are mostly consistent with the SF region of the O1-BPT diagram, given their upper limits. Several of these limits also lie above the \citet{Kewley2001} maximum starburst boundary, so contributions from shock-heated gas in the ISM cannot be ruled out for these galaxies \citep{Sutherland2017}. The stacked CECILIA point lies close to the main SF locus but shows a slight offset toward higher O1, while the faintest galaxies (with only upper limits) lie below the sequence.

\citet{Clarke2023} found elevated O1 at fixed O3 relative to local H~II regions, arguing that this offset is consistent with ionization by $\alpha$-enhanced massive stars, as expected for rapidly-forming galaxies in the early Universe. The published targets in their sample have typical stellar masses of $M_* \sim 10^9 \ M_{\odot}$ and star formation rates of $\sim 5$--$100 \ M_{\odot}$/yr, indicating that our sample is less evolved and likely represents the least mature sample of $z\sim2-3$ galaxies with constraints on [O~I] emission to date. Despite the potential differences in population properties between our sample and those of \citet{Clarke2023} and \citet{Cameron2023}, all three samples seem to occupy a similar space in the O1 BPT diagram or have similar upper-limits.

\section{Discussion}

\subsection{Comparison to Photoionization Models}
\label{Photoionization}

\begin{figure*}[h]
   \centering
   \includegraphics[width=0.85\textwidth]{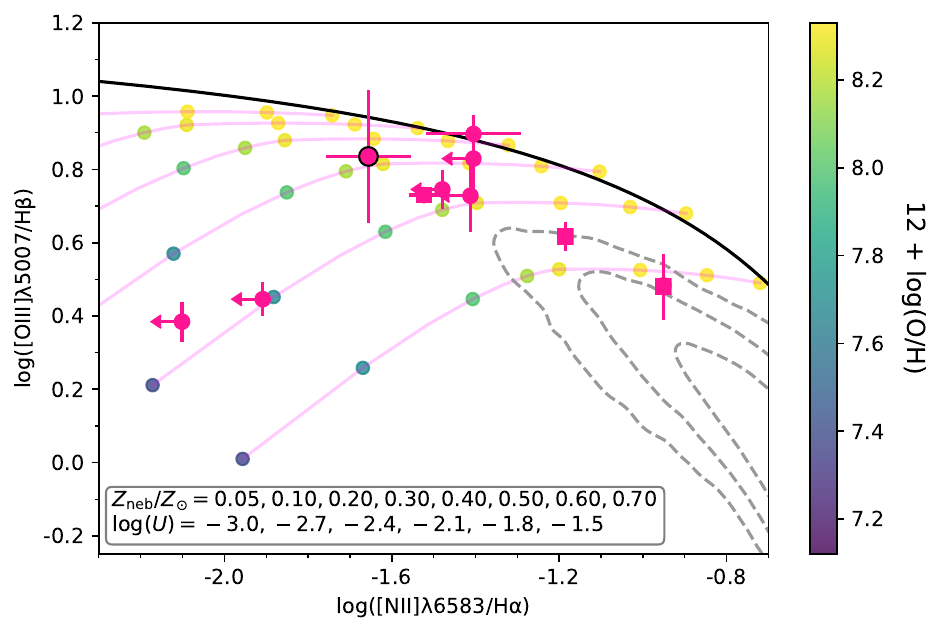}
   \caption{The N2-BPT predictions from the \texttt{Cloudy} photoionization models are shown, using input spectra from BPASS stellar populations. The pink lines represent tracks of fixed $\log U$, where higher $\log U$ values result in higher O3 ratios. The model points are color-coded by O/H as indicated by the color bar. We overplot the CECILIA narrowband sample (pink points), fLBG sources (pink squares), and SDSS~$z \sim 0$ galaxies (gray contours).}
   \label{fig:all_bpt_diagrams}
\end{figure*}

\begin{figure*}[b]
   \centering
   \includegraphics[width=0.48\textwidth]{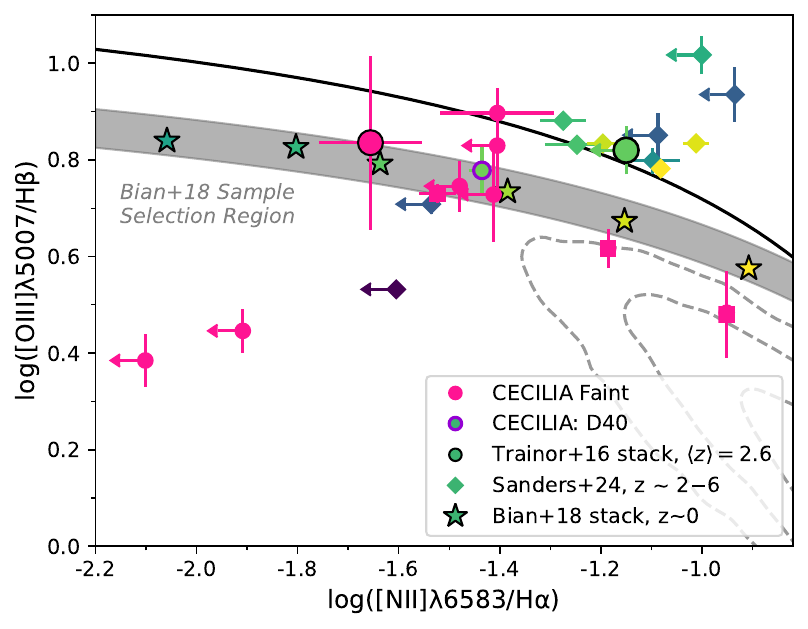}
    \hskip -1ex
   \includegraphics[width=0.52\textwidth]{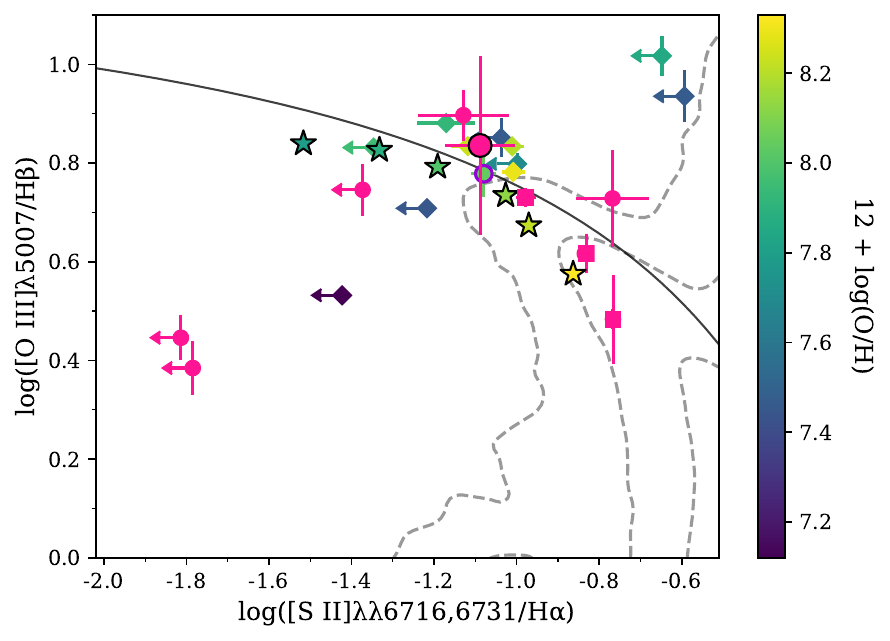}
   \caption{N2 (\textit{left}) and S2-BPT (\textit{right}) diagrams. Our data points are compared to 11 galaxies at $z \sim 2$--6 from \citet{Sanders2024} (diamond markers) and the stacked points from \citet{Bian2018} (stars), both color-coded by their direct metallicities, as indicated by the colorbar on the right. The \citet{Bian2018} sample was selected on the basis of N2-BPT line ratios, with the gray shading denoting the selection region. In the N2-BPT diagram, the stacked point from \citetalias{Trainor2016} and D40 from \citetalias{2024ApJ...964L..12R} (black and purple outlines, respectively) are shown as circles with the same O/H-based central color coding.}
   \label{fig:Sanders_Comparison}
\end{figure*} 

To interpret the observed decline in O3 at low values of N2 or S2, we compare our data to photoionization models. Figure~\ref{fig:all_bpt_diagrams} presents our theoretical stellar population synthesis and photoionization model grid for SF galaxies. For consistency with \citetalias{Trainor2016} and \citet{stro2018}, we use models generated using \textsc{Cloudy} v13.02. \citep{ferl2013} with BPASS stellar populations \citep{Stanway2018} as the input spectrum, assuming a stellar metallicity of $Z_* = 0.001$. These models span a range of ionization parameters from $\log U = -3.0$ to $-1.5$ and a nebular metallicity $Z_{\mathrm{neb}}/Z_{\odot}$ between 0.05 and 0.70, capturing the diversity in nebular properties seen in high-redshift SF galaxies. The gas density in these photoionization models is fixed at $n_{\mathrm{H}} = 300$~cm$^{-3}$, consistent with electron density measurements in $\langle z \rangle = 2.3$ KBSS galaxies (\citealt{stei2016}; \citetalias{stro2017})\footnote{Assuming a lower density $n_{\mathrm{H}} \sim 100$--$200$~cm$^{-3}$ shifts the model curves slightly to lower O3, but this does not change the interpretation of our results.}.

Following analogous methods in \citet{stei2016}, we apply corrections to the model points to account for expected variation in elemental abundance patterns for rapidly-forming galaxies at high redshift. 
Specifically, for the N2-BPT diagram, we apply a metallicity-dependent scaling to the [NII] fluxes, adjusting the log(N/O) ratio according to the relation $\log(\text{N/O}) = 1.64 \log(\text{O/H}) - 0.86$, which reflects the secondary increase in N/O with respect to O/H \citepalias{stro2017}, with a lower limit of $\log(\text{N/O}) = -1.5$.


From the N2-BPT diagram in Figure~\ref{fig:all_bpt_diagrams}, most CECILIA-faint galaxies fall in the range $\log U \sim -2.7$ to $-2.4$, with fC23 at lower ionization, $\log U \sim -3.0$. The majority of points with N2 detections correspond to nebular metallicities of $Z_{\mathrm{neb}}/Z_{\odot} \sim 0.2$--$0.5$, or $12 + \log(\text{O/H}) \sim 8.0$--$8.4$, as indicated by both individual galaxies and the stacked point. This location also marks the point in the model grid where O3 reaches its peak: below $12 + \log(\mathrm{O/H}) \sim 8.0$ ($Z_{\mathrm{neb}}/Z_{\odot} \lesssim 0.2$), the O3 ratio begins to decrease with further declines in oxygen abundance due to the sharp drop in [O~III] emissivity at extremely low metallicities. In contrast, the two faintest galaxies, NB2089 and NB2875, show higher ionization ($\log U > -2.7$) and lower metallicity ($Z_{\mathrm{neb}}/Z_{\odot} < 0.1$, or $12 + \log(\text{O/H}) \lesssim 7.6$). Many data points represent upper limits on the N2 ratio, implying lower limits on $\log U$ and upper limits on $Z_{\mathrm{neb}}/Z_{\odot}$.


\subsection{Comparison to T$_e$-based Metallicity Samples}
\label{sec:comparison_to_direct}
Due to the uncertain correspondence between our galaxies and the assumptions of photoionization models, we also compare to empirical O/H measurements from the literature. Figure~\ref{fig:Sanders_Comparison} displays the CECILIA Faint points in the N2 and S2-BPT diagrams along with several samples of galaxies at $z>2$ with direct $T_e$-based measurements of O/H: the stacked KBSS-Ly$\alpha$ sample of 60 faint LAEs from \citetalias{Trainor2016}, the CECILIA galaxy D40 from \citetalias{2024ApJ...964L..12R}, and the sample of 46 galaxies at at $z = 2.1-8.7$ from the CEERS survey and other JWST/NIRSpec observations from the literature compiled by \citet{Sanders2024}. We also compare to a $z\sim0$ sample of ``high-redshift analogs'' with stacked $T_e$ measurements from \citet{Bian2018}.

In general, the CECILIA Faint stack and several of the individual sources lie near the \citetalias{Trainor2016} and D40 points in the N2-BPT diagram, and near  D40\footnote{The KBSS-Ly$\alpha$ stack spectrum from \citetalias{Trainor2016} does not extend sufficiently far redward to cover the [\ion{S}{2}]$\lambda\lambda$6717,6731 doublet.} in the S2-BPT diagram. \citetalias{2024ApJ...964L..12R} measured $12 + \log(\text{O/H}) = 8.07\pm0.06$ for D40, and \citetalias{Trainor2016} similarly find $12 + \log(\text{O/H}) = 8.04\pm0.19$ for the KBSS-Ly$\alpha$ stack, although they suggest that individual sources have metallicities as low as $12 + \log(\text{O/H}) \sim 7.0$--$7.5$ using similar O3-based arguments to those discussed in this work.

The \citet{Sanders2024} sample spans $12 + \log(\text{O/H}) = 7.12 \pm 0.11$ to $8.33 \pm 0.14$. Again, we find that the points that lie closest to the CECILIA Faint stack in in Figure~\ref{fig:Sanders_Comparison} have typical abundances $12 + \log(\text{O/H})\approx 8.0$, particularly in the S2-NPT diagram where the \citet{Sanders2024} points lie close to D40 and the CECILIA Faint stack.  In conjunction with the photoionization model comparison in Sec.~\ref{Photoionization},  an abundance $12 + \log(\text{O/H})\approx 8.0$ ($Z_{\mathrm{neb}} \approx 0.2\,\textrm{Z}_{\odot}$) thus appears typical of the CECILIA Faint sample.

NB2089 and NB2875 exhibit lower O3 ratios and lower N2 and S2 limits than even the most metal-poor galaxy in the \citet{Sanders2024} sample, which has $12 + \log(\text{O/H}) = 7.12 \pm 0.11$. This comparison, together with the trends of the models shown in Figure~\ref{fig:all_bpt_diagrams}, suggest that NB2089 and NB2875 are likely to have even lower metallicities.

These faintest and presumably lowest-O/H galaxies also depart significantly from the high-redshift analog galaxy stacks of \citet{Bian2018}. The galaxies in that sample were selected for elevated O3 ratios similar to those in more massive KBSS and MOSDEF galaxies, which are known to be offset from the SF sequence established at $z\sim0$ as previously discussed. As seen in Figure \ref{fig:Sanders_Comparison}, these analogs indeed track high-redshift galaxies at intermediate values of N2 and S2: they closely match the CECILIA Faint stack near $12+\log(\textrm{O/H})\approx8.0$, as well as D40 and many of the other $z>2$ points. However, the \citet{Bian2018} track diverges at lower values of S2 and N2 where our apparent low-O/H galaxies fall outside their selection. This divergence is in fact definitional, since their selection criteria (shown in the gray shaded region of Figure~\ref{fig:Sanders_Comparison}) 
systematically excludes low-O3 systems like NB2089, NB2875, and the lowest-O/H object from \citet{Sanders2024}. Thus, while informative for identifying analogs of more massive $z \sim 2$ galaxies, we caution that the \citet{Bian2018} galaxies and other similarly-selected samples likely do not represent the full diversity of the faint, high-redshift galaxy population, particularly at the lowest metallicities. 

Finally, Figure~\ref{fig:Sanders_Comparison} also suggests an intriguing variation in the [\ion{S}{2}] and/or [\ion{N}{2}] emission between the CECILIA and \citet{Sanders2024} samples. While both D40 and the CECILIA Faint galaxies track the \citet{Sanders2024} galaxies in S2 space, they are significantly shifted to lower N2 in the left panel --- that is, there are no galaxies in the published CECILIA sample\footnote{Note that the \citetalias{Trainor2016} stack has a relatively weak upper limit on N2, making it consistent with both the \citet{Sanders2024} points at $12 + \log(\text{O/H}) \sim 8$ and the CECILIA-faint sources.
} that have direct analogs in the \citet{Sanders2024} sample across both diagrams. This contrast suggests differences in ionization conditions, abundance ratios, or other ISM properties between the CECILIA sample and those compiled by \citet{Sanders2024}.

\citet{Shapley2019} found that galaxies with higher star formation rate surface densities ($\Sigma_{\mathrm{SFR}}$) tend to show lower S2 ratios at fixed O3, likely due to a decreasing fraction of Balmer flux contributed by diffuse ionized gas (DIG) as $\Sigma_{\mathrm{SFR}}$ increases. This DIG effect may partially account for the N2 and S2 differences observed across these samples. However, the interpretation of S2 is further complicated by the fact that S$^+$ traces lower-ionization zones outside of classical H II regions — due to the relatively low ionization potential of S$^0$ ($\sim10.4$ eV) — unlike N$^+$, which more closely traces H II regions. As a result, S2 may respond differently to variations in the ionizing spectrum.

Lastly, recent work including \citetalias{2024ApJ...964L..12R} has found variation in S/O with respect to the solar ratio, perhaps due to the contribution of Type Ia supernovae to S production at late times, while N/O also varies with O/H and star-formation history (e.g., \citealt{masters2016,matthee2018}; \citetalias{stro2017}; \citealt{Strom2022}). Given the many factors in play, further analysis of the variation in N2 and S2 at high redshift is beyond the scope of this work---future progress may be facilitated by larger samples of galaxies with secure abundances via the auroral [\ion{N}{2}] $\lambda5755$ and [\ion{S}{3}] $\lambda6314$ emission lines.

\subsection{Ly$\alpha$ Strength as a Probe of Metallicity}
\label{lya_discussion}

Previous studies have established a strong link between Ly$\alpha$ emission and nebular properties, particularly gas-phase metallicity and ionization state. \citetalias{Trainor2016} found that galaxies with stronger Ly$\alpha$ emission tend to occupy the upper-left region of the N2-BPT diagram, characterized by high O3 and low N2 values. This trend suggests that high Ly$\alpha$ EWs correlate with lower metallicities, as reduced metal content produces harder stellar spectra and reduced dust attenuation to enhance Ly$\alpha$ production and escape (e.g., see extended discussion by \citealt{Trainor2019}). 
In general, higher Ly$\alpha$ EWs correlate with higher O3 ratios for metallicities above approximately 10\% Z$_\odot$ — i.e., over the range of metallicities where O3 is inversely correlated with O/H (\citetalias{Trainor2016}; \citealt{Cullen2020, Du2021}).
Similarly, \citet{Maseda2023} confirmed that galaxies with the highest Ly$\alpha$ EWs exhibit systematically lower gas-phase metallicities, with some reaching values as low as $\sim$1.5\% Z$_\odot$. \citet{Maseda2023} reported that this relation holds even as O3 begins to decline at extremely low metallicities, suggesting that a more fundamental trend between Ly$\alpha$ and oxygen abundance drives the apparent association with the O3 ratio.

\begin{figure}[h]
   \centering
   \includegraphics[width=0.47\textwidth]{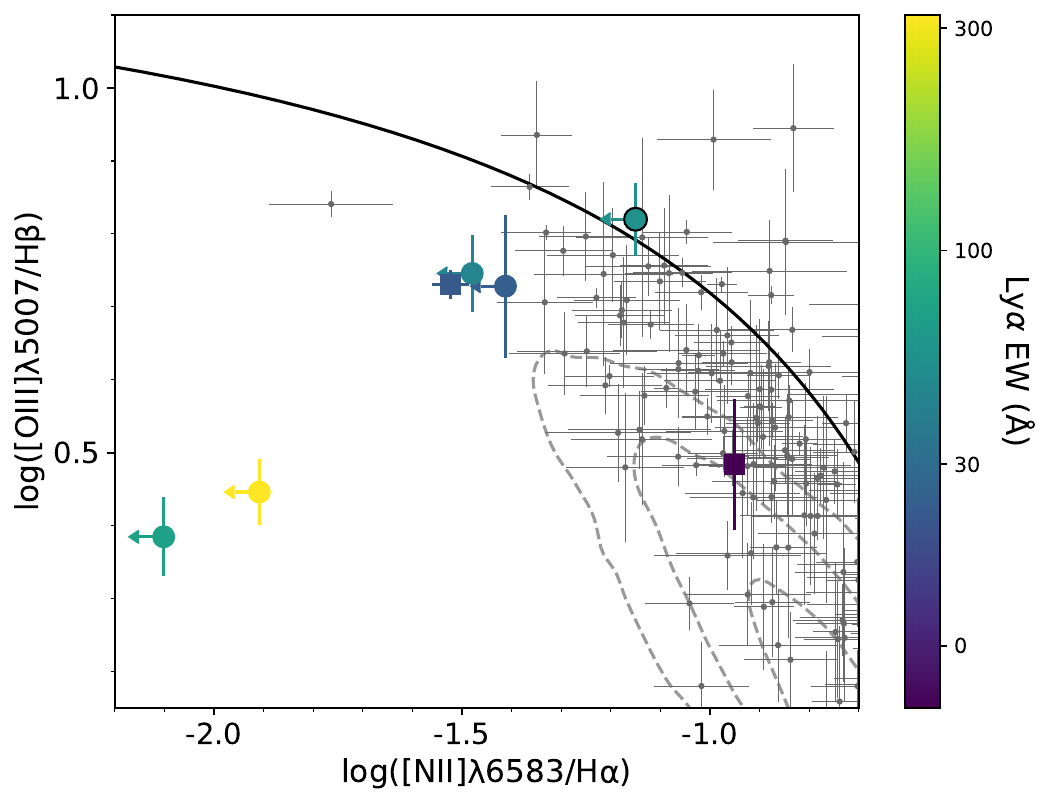}
   \caption{N2-BPT diagram with the same features as Figure~\ref{fig:n2bpt}, but excluding all stacked points and galaxies without reported Ly$\alpha$ EWs. CECILIA faint sources are shown as circles for narrow-band detections and squares for fLGBs. The \citetalias{Trainor2016} stack point is plotted as a circle with a black outline. All points are color-coded by their Ly$\alpha$ EWs.}
   \label{fig:lya_plot}
\end{figure}

Our sample follows a similar pattern, as shown in Figure \ref{fig:lya_plot}. Based on Ly$\alpha$ EW measurements from \citetalias{Trainor2016} and \citet{Trainor2019}, the two galaxies with the highest Ly$\alpha$ EWs exhibit the lowest O3 values and N2 limits, supporting the hypothesis that metallicity plays a dominant role in regulating Ly$\alpha$ emission. 
In contrast, the CECILIA fLBGs have the highest N2 values and the lowest Ly$\alpha$ EWs in our sample.

It is important to note that Ly$\alpha$ emission is highly sensitive to additional factors beyond metallicity, such as gas kinematics and the structure of the interstellar and circumgalactic media. The observed scatter in Ly$\alpha$ emission across different samples underscores the fact that while lower metallicity facilitates Ly$\alpha$ production and escape, variations in neutral gas covering fraction, outflows, and ISM geometry play a crucial role in shaping the net Ly$\alpha$ output of galaxies. The intensity and hardness of the radiation field, as well as dust content, are themselves correlated with metallicity, whereas gas kinematics and structure are less directly linked to metallicity and primarily affect Ly$\alpha$ photon escape. For further discussion of empirical correlations relating Ly$\alpha$ production and escape, see \citet{Trainor2019}, \citet{Runnholm2020}, and \citet{Du2021}.

\subsection{EoR analogs}
\label{sec:eor_analogs}

The study of LAEs at redshift $z \sim 2-3$ provides key insights into the ionization conditions and ISM properties of galaxies during the EoR. Continuum-faint and Ly$\alpha$-emitting galaxies are known to have properties that facilitate the escape of Lyman continuum (LyC) photons \citep{trai2015, verh2015, Dijkstra2016, stei2018}. By investigating the physical conditions of LAEs at these redshifts, we can better understand the mechanisms that regulate Ly$\alpha$ escape, ionization states, and potential LyC photon leakage — factors that are essential in determining the contribution of SF galaxies to the reionization process.

Previous studies have shown that strong O3 and O32\footnote{O32 $\equiv \log$([\ion{O}{3}]$\lambda\lambda4959,5007$/[\ion{O}{2}]$\lambda\lambda$3726,3729)} ratios are associated with the production of ionizing and Ly$\alpha$ photons (e.g., \citetalias{Trainor2016}, \citealt{Trainor2019}) and tend to correlate with LyC leakage, such that these ratios have been used to identify LyC-emitting galaxies \citep[e.g.,][]{deBarros2016, izot2016a, izot2018a, Fletcher2019}. 
However, as demonstrated in this work and by \citet{Maseda2023}, this correlation is likely driven by low metallicity rather than the O3 ratio itself. This has important implications for studies of reionization-era galaxies and their lower-redshift analogs, where selecting galaxies based on high O3 ratios may inadvertently exclude the lowest-metallicity systems, which are potentially the strongest LyC and Ly$\alpha$ emitters. 


As metallicity decreases with increasing redshift on average, identifying extremely metal-poor galaxies remains a key objective. Although O3-based selection has proven effective for identifying LyC emitters at moderate metallicities, its double-valued behavior at low metallicity may limit its utility in the most metal-poor regime. In these cases, alternative or complementary diagnostics such as N2 or S2 may provide more reliable selection criteria. As described in Sec.~\ref{sec:comparison_to_direct}, even criteria intended to select high-redshift analogs more generally --- not only LyC emitters --- such as the \citet{Bian2018} selection may select against the lowest-metallicity systems with lower O3 ratios.

In addition to the N2 and S2 ratios, O32-based diagnostics\footnote{e.g., the R23-O32 diagram, which uses both O32 and $\textrm{R23}\equiv \log($[\ion{O}{2}]$\lambda\lambda3726,3729/\textrm{H}\beta)+\log($[\ion{O}{3}]$\lambda\lambda4959,5007/\textrm{H}\beta$)  to break the double-valued nature of both R23 and O3.} may offer practical advantages for high-redshift observations due to the shorter rest-frame wavelengths of their constituent lines compared to N2 and S2. Future searches for strong LyC emitters and metal-poor galaxies should incorporate complementary line ratios to mitigate selection biases and accurately probe the low-metallicity regime.

\section{Conclusions}
In this paper, we have presented ultra-deep rest-optical spectra of continuum-faint galaxies at $z \sim 2.5$, obtained as part of the CECILIA program. We analyzed a sample of nine continuum-faint galaxies observed with JWST/NIRSpec, including narrowband-selected LAEs, fLBGs, and serendipitous detections. These galaxies represent some of the lowest-mass and faintest star-forming galaxies spectroscopically observed at $z \sim 2-3$. Our analysis explores their dust extinction, SFRs, electron densities, ionization conditions, and nebular properties. The key findings of our study are summarized below:
\begin{enumerate}
    
    \item We found the dust extinction of our sample varying from \( E(B-V) = 0.05 \) to \( 0.95 \) using the H$\alpha$/H$\beta$ ratio, with a median of \( 0.23 \), slightly higher than the median values of the KBSS-Ly$\alpha$ parent sample \citepalias{Trainor2016}. The dust content of our four LAEs is consistent with values reported in previous studies at similar redshifts. 
    \textbf{Figure
    \ref{fig:ebv_properties}};
    \textbf{Section \ref{section_dust_attenuation}}.
    
    \item The sample exhibits a range of SFRs from 0.63 to 5.43 M$_{\odot}$ yr$^{-1}$, with a median of 1.54 M$_{\odot}$ yr$^{-1}$, which is lower than the median values of the KBSS-Ly$\alpha$ sample and the KBSS LBG sample \citepalias{stro2017}. This is due to the targeted galaxies being at the faint end of the population by every metric. 
    \textbf{Figure
    \ref{fig:ebv_properties}};
    \textbf{Section \ref{star_formation_rates_section}}.

    \item Electron densities were estimated using the [SII]$\lambda6717/\lambda6731$ ratio for two galaxies with sufficient S/N, yielding values of $n_e = 205^{+39}_{-55}$~cm$^{-3}$ and $n_e = 95^{+93}_{-48}$~cm$^{-3}$. These are slightly below the typical electron densities observed in $z \sim 2$ galaxies. \textbf{Section \ref{sec:electron_densities}}.

    \item We constructed the N2, S2, and O1 BPT diagrams to compare our CECILIA faint galaxies with other similar and higher redshift SF galaxies, as well as with local galaxies. Most of our sample exhibits low N2 and high O3 values, indicative of low metallicity and extreme ionization conditions. Two sources with the lowest N2 upper limits exhibit lower O3 values, likely due to very low oxygen abundances (12 + $\log(\text{O/H}) \ll 8$). This paper also presents the first analysis of [O~I] emission in continuum-faint $z\sim2-3$ galaxies. \textbf{Figures \ref{fig:n2bpt}, \ref{fig:s2bpt}, \ref{fig:o1bpt}}; \textbf{Section \ref{Diagnostic_section}}.

    \item We compared our CECILIA-faint galaxies to photoionization models and galaxies with $T_e$-based metallicity measurements to estimate their nebular properties. The galaxies show a range of ionization parameters with \( \log U = -2.7 \) to \( -2.4 \) and typical metallicities of \( Z_{\mathrm{neb}}/Z_{\odot} \sim 0.2 - 0.5 \), consistent with \citetalias{Trainor2016} and \citetalias{2024ApJ...964L..12R}. Two galaxies in the sample fall below this range, exhibiting lower metallicities of \( Z_{\mathrm{neb}}/Z_{\odot} < 0.1 \). Comparison to SF galaxies at redshifts \( z = 2.1-8.7 \) from \citet{Sanders2024} likewise suggests a similar metallicity. However, our typical CECILIA faint galaxies exhibit differences in S2 and/or N2 ratios compared to the \citet{Sanders2024} sample, suggesting differences in ionization conditions, ISM properties, or observational effects.
    \textbf{Figures \ref{fig:all_bpt_diagrams}, \ref{fig:Sanders_Comparison}}; \textbf{Sections \ref{Photoionization}, \ref{sec:comparison_to_direct}}.

    \item Comparisons to local high-redshift analogs show that samples selected based on elevated O3 ratios agree well with the intermediate-metallicity points, but fail to capture the downturn observed in the apparently very low-metallicity sources and will therefore miss them. \textbf{Figure~\ref{fig:Sanders_Comparison}; Section~\ref{sec:comparison_to_direct}}.

    \item The two galaxies with low O3 values noted above also have the highest Ly$\alpha$ EWs in our sample, supporting the hypothesis that metallicity plays an important role in regulating Ly$\alpha$ emission. This finding is consistent with previous studies showing that galaxies with stronger Ly$\alpha$ emission tend to have lower gas-phase metallicities and higher ionization states. \textbf{Figure \ref{fig:lya_plot}}; \textbf{Section \ref{lya_discussion}}.

    \item Our study of LAEs at redshift $z \sim 2-3$ provides new insights into the ionization conditions and ISM properties during the EoR. We highlight the role of continuum-faint galaxies in Ly$\alpha$ escape and LyC photon leakage. This finding underscores the importance of identifying low-metallicity galaxies, which may be missed if selection relies solely on O3 ratios. We recommend using complementary diagnostics, such as N2 and O32, to more effectively identify these galaxies, at high redshifts. \textbf{Section \ref{sec:eor_analogs}}.
\end{enumerate}


Future work will focus on quantifying the metallicities of these galaxies through more detailed modeling and strong-line indicators, as well as exploring the relationships between metallicity and other global properties including their physical sizes, masses, ionization conditions, and star formation rates. 

Moreover, upcoming JWST observations will facilitate the study of more faint SF galaxies, greatly enhancing our understanding of the early universe. These advancements will provide deeper insights into the physical processes at play in low-mass, faint galaxies, and offer new perspectives on the role these galaxies played in cosmic reionization and the evolution of the first galaxies.

\begin{acknowledgments}
R.F.T, A.L.S., and G.C.R. acknowledge partial support from the JWST-GO-02593.008-A, JWST-GO-02593.004-A, and JWST-GO-02593.006-A grants, respectively.
N.S.J.R. was also supported by JWST-GO-02593.004-A. These funds were provided by NASA through a grant from the Space Telescope Science Institute, which is operated by the Association of Universities for Research in Astronomy, Inc., under NASA contract NAS 5-03127.
R.F.T. also acknowledges support from the Pittsburgh Foundation (grant ID UN2021-121482) and the Research Corporation for Scientific Advancement (Cottrell Scholar Award, grant ID 28289). M.V.M. is supported by the National Science Foundation via AAG grant 2205519.

This work is primarily based on observations made with NASA/ESA/CSA JWST, associated with PID 2593, which can be accessed via doi:\dataset[10.17909/x66z-p144]{https://doi.org/10.17909/x66z-p144}. The data were obtained from the Mikulski Archive for Space Telescopes (MAST) at the Space Telescope Science Institute, which is operated by the Association of Universities for Research in Astronomy, Inc., under NASA contract NAS 5-03127 for JWST.

\end{acknowledgments}

\textit{Facilities:} JWST (NIRSpec)


\textit{Software:} \textsc{matplotlib} \citep{hunt2007}, \textsc{NumPy} \citep{harr2020}, \textsc{SciPy} \citep{virt2020}, \textsc{calwebb} \citep{calwebb_v1.12.5_2023}, \textsc{grizli} \citep{griz2023}, \textsc{msaexp} \citep{msaexp2022}, \textsc{NSClean} \citep{raus2023}.

\appendix


\section{Individual G235M spectra for serendipitous sources}
In this Appendix section, we present the individual G235M spectra for the two serendipitous sources C31b and BX587b, along with their fits and 2D spectrograms. The meaning of their legend is the same as in Figure \ref{fig:1D spectrum}.







\begin{figure*}[h]
   \centering
   \includegraphics[width=0.96\textwidth]{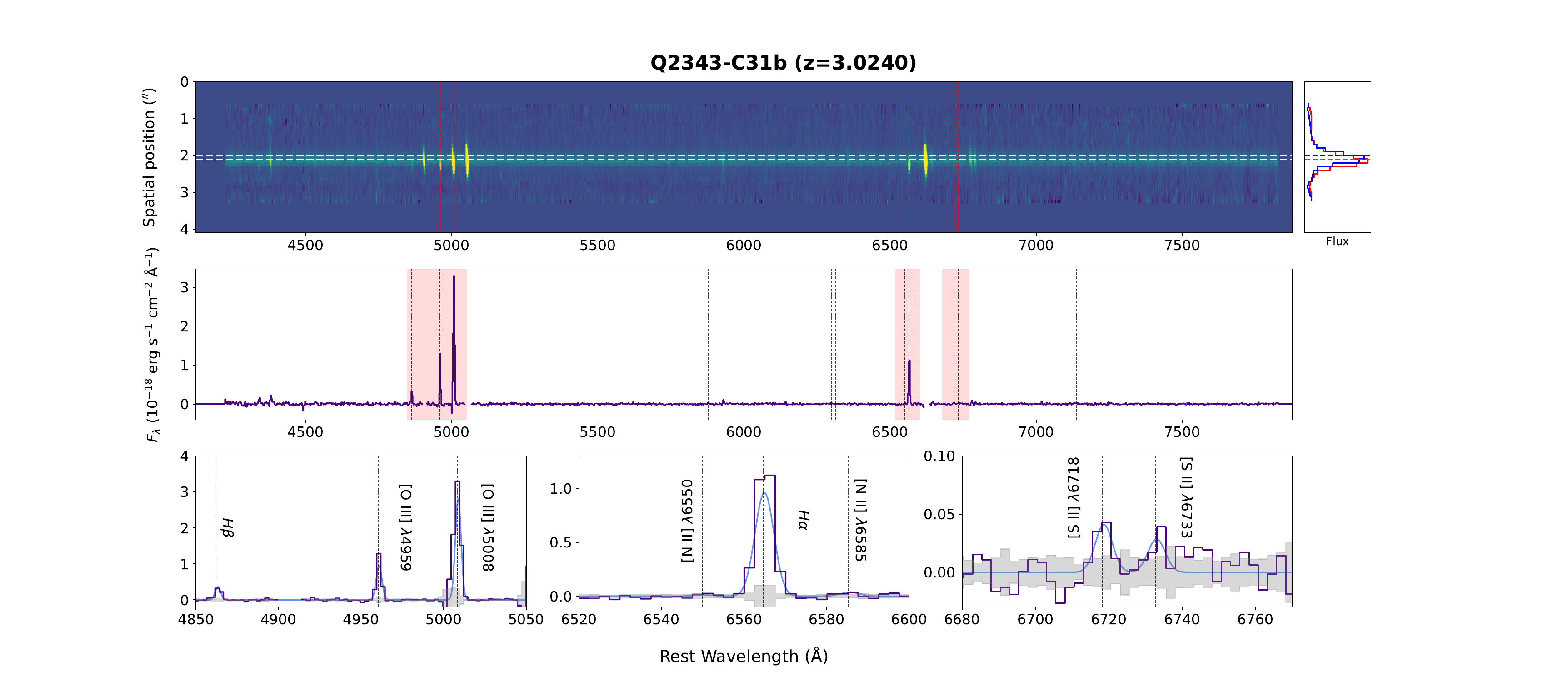}
   \caption{The rest-frame NIRSpec spectrum of the serendipitous source Q2343-C31b. 
\textit{Top:} 2D spectrogram of the source, with the emission lines marked by red vertical lines and the two white dashed lines indicating the centers of the main and serendipitous sources. Adjacent to the spectrogram, the panel shows the 1D spatial distribution of the continuum for the main source in blue (with the blue dashed line marking the center of the main source). Additionally, the 1D spatial distribution of the brightest emission lines from the serendipitous source is shown in red (with the red dashed line marking the center of the serendipitous source). 
\textit{Middle:} Full spectral range of the G235M spectrum for the source. Black dashed vertical lines and shaded bands indicate the positions of key spectral features — most of which are shown in the bottom panels — and emission lines with $>2\sigma$ detections. The 1D extracted spectrum is shown in purple. 
\textit{Bottom:} Zoom-in panels of key spectral features. The Gaussian+linear continuum fit to the emission lines is shown in light blue, and the error region is shown in gray.
}
   \label{fig:C31_ser}
\end{figure*}

\begin{figure*}[h]
   \centering
   \includegraphics[width=0.96\textwidth]{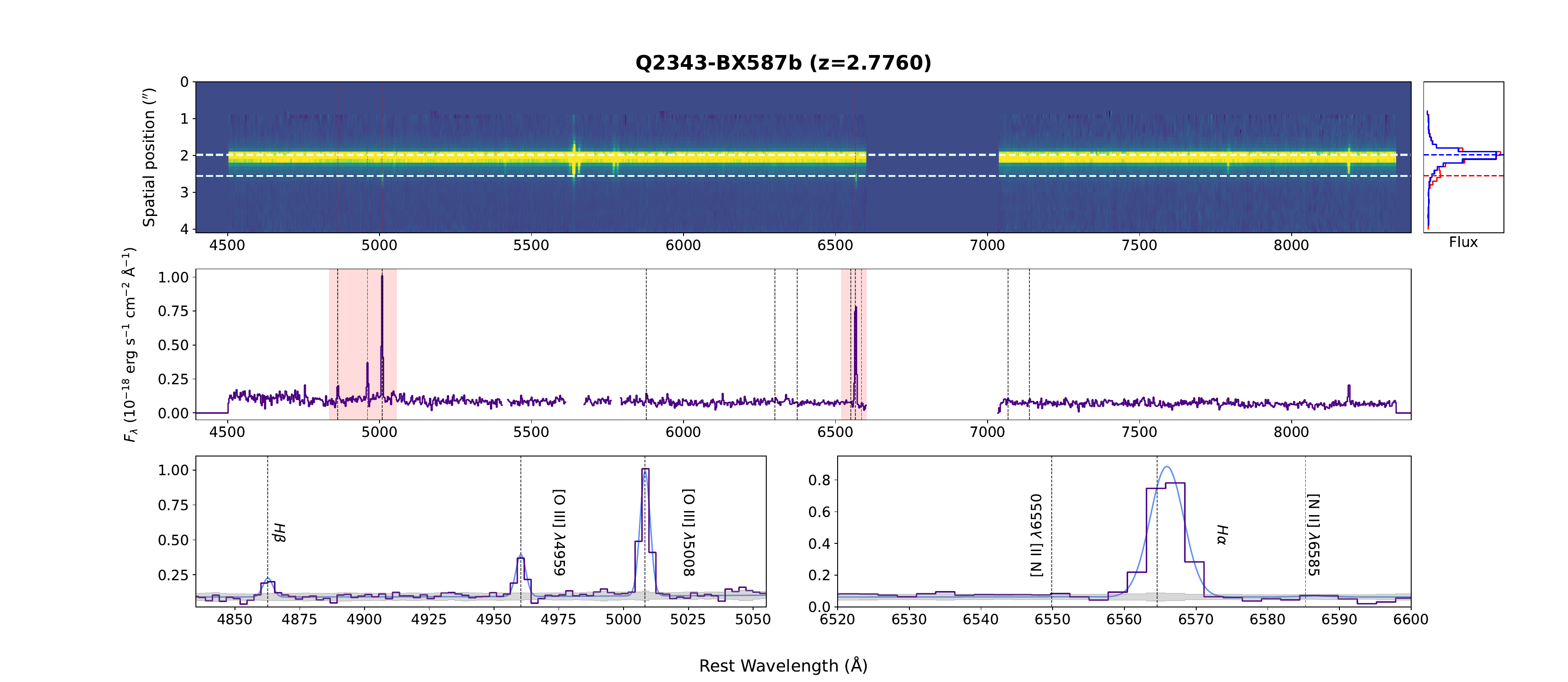}
   \caption{The rest-frame NIRSpec spectrum of the serendipitous source Q2343-BX587b. Legend names are the same as in Figure \ref{fig:C31_ser}.}
   \label{fig:BX587_ser}
\end{figure*}

\bibliographystyle{aasjournal}
\bibliography{aggregate_refs}

\begin{thebibliography}{}
\expandafter\ifx\csname natexlab\endcsname\relax\def\natexlab#1{#1}\fi
\providecommand{\url}[1]{\href{#1}{#1}}
\providecommand{\dodoi}[1]{doi:~\href{http://doi.org/#1}{\nolinkurl{#1}}}
\providecommand{\doeprint}[1]{\href{http://ascl.net/#1}{\nolinkurl{http://ascl.net/#1}}}
\providecommand{\doarXiv}[1]{\href{https://arxiv.org/abs/#1}{\nolinkurl{https://arxiv.org/abs/#1}}}

\bibitem[{{Abazajian} {et~al.}(2009){Abazajian}, {Adelman-McCarthy}, {Ag{\"u}eros}, {Allam}, {Allende Prieto}, {An}, {Anderson}, {Anderson}, {Annis}, {Bahcall}, {Bailer-Jones}, {Barentine}, {Bassett}, {Becker}, {Beers}, {Bell}, {Belokurov}, {Berlind}, {Berman}, {Bernardi}, {Bickerton}, {Bizyaev}, {Blakeslee}, {Blanton}, {Bochanski}, {Boroski}, {Brewington}, {Brinchmann}, {Brinkmann}, {Brunner}, {Budav{\'a}ri}, {Carey}, {Carliles}, {Carr}, {Castander}, {Cinabro}, {Connolly}, {Csabai}, {Cunha}, {Czarapata}, {Davenport}, {de Haas}, {Dilday}, {Doi}, {Eisenstein}, {Evans}, {Evans}, {Fan}, {Friedman}, {Frieman}, {Fukugita}, {G{\"a}nsicke}, {Gates}, {Gillespie}, {Gilmore}, {Gonzalez}, {Gonzalez}, {Grebel}, {Gunn}, {Gy{\"o}ry}, {Hall}, {Harding}, {Harris}, {Harvanek}, {Hawley}, {Hayes}, {Heckman}, {Hendry}, {Hennessy}, {Hindsley}, {Hoblitt}, {Hogan}, {Hogg}, {Holtzman}, {Hyde}, {Ichikawa}, {Ichikawa}, {Im}, {Ivezi{\'c}}, {Jester}, {Jiang}, {Johnson}, {Jorgensen}, {Juri{\'c}}, {Kent}, {Kessler}, {Kleinman}, {Knapp},
  {Konishi}, {Kron}, {Krzesinski}, {Kuropatkin}, {Lampeitl}, {Lebedeva}, {Lee}, {Lee}, {French Leger}, {L{\'e}pine}, {Li}, {Lima}, {Lin}, {Long}, {Loomis}, {Loveday}, {Lupton}, {Magnier}, {Malanushenko}, {Malanushenko}, {Mandelbaum}, {Margon}, {Marriner}, {Mart{\'\i}nez-Delgado}, {Matsubara}, {McGehee}, {McKay}, {Meiksin}, {Morrison}, {Mullally}, {Munn}, {Murphy}, {Nash}, {Nebot}, {Neilsen}, {Newberg}, {Newman}, {Nichol}, {Nicinski}, {Nieto-Santisteban}, {Nitta}, {Okamura}, {Oravetz}, {Ostriker}, {Owen}, {Padmanabhan}, {Pan}, {Park}, {Pauls}, {Peoples}, {Percival}, {Pier}, {Pope}, {Pourbaix}, {Price}, {Purger}, {Quinn}, {Raddick}, {Re Fiorentin}, {Richards}, {Richmond}, {Riess}, {Rix}, {Rockosi}, {Sako}, {Schlegel}, {Schneider}, {Scholz}, {Schreiber}, {Schwope}, {Seljak}, {Sesar}, {Sheldon}, {Shimasaku}, {Sibley}, {Simmons}, {Sivarani}, {Allyn Smith}, {Smith}, {Smol{\v{c}}i{\'c}}, {Snedden}, {Stebbins}, {Steinmetz}, {Stoughton}, {Strauss}, {SubbaRao}, {Suto}, {Szalay}, {Szapudi}, {Szkody}, {Tanaka},
  {Tegmark}, {Teodoro}, {Thakar}, {Tremonti}, {Tucker}, {Uomoto}, {Vanden Berk}, {Vandenberg}, {Vidrih}, {Vogeley}, {Voges}, {Vogt}, {Wadadekar}, {Watters}, {Weinberg}, {West}, {White}, {Wilhite}, {Wonders}, {Yanny}, {Yocum}, {York}, {Zehavi}, {Zibetti}, \& {Zucker}}]{Abazajian2009}
{Abazajian}, K.~N., {Adelman-McCarthy}, J.~K., {Ag{\"u}eros}, M.~A., {et~al.} 2009, \apjs, 182, 543, \dodoi{10.1088/0067-0049/182/2/543}

\bibitem[{{Andrews} \& {Martini}(2013)}]{Andrews2013}
{Andrews}, B.~H., \& {Martini}, P. 2013, \apj, 765, 140, \dodoi{10.1088/0004-637X/765/2/140}

\bibitem[{{Asplund} {et~al.}(2021){Asplund}, {Amarsi}, \& {Grevesse}}]{aspl2021}
{Asplund}, M., {Amarsi}, A.~M., \& {Grevesse}, N. 2021, \aap, 653, A141, \dodoi{10.1051/0004-6361/202140445}

\bibitem[{{Baldwin} {et~al.}(1981){Baldwin}, {Phillips}, \& {Terlevich}}]{Baldwin1981}
{Baldwin}, J.~A., {Phillips}, M.~M., \& {Terlevich}, R. 1981, \pasp, 93, 5, \dodoi{10.1086/130766}

\bibitem[{{Bian} {et~al.}(2018){Bian}, {Kewley}, \& {Dopita}}]{Bian2018}
{Bian}, F., {Kewley}, L.~J., \& {Dopita}, M.~A. 2018, \apj, 859, 175, \dodoi{10.3847/1538-4357/aabd74}

\bibitem[{{Brammer}(2022{\natexlab{a}})}]{msaexp}
{Brammer}, G. 2022{\natexlab{a}}, {msaexp: NIRSpec analyis tools}, 0.3.4, Zenodo,  Zenodo, \dodoi{10.5281/zenodo.7299500}

\bibitem[{{Brammer}(2022{\natexlab{b}})}]{msaexp2022}
---. 2022{\natexlab{b}}, {msaexp: NIRSpec analyis tools}, 0.3.4, Zenodo,  Zenodo, \dodoi{10.5281/zenodo.7313329}

\bibitem[{{Brammer}(2023{\natexlab{a}})}]{grizli}
---. 2023{\natexlab{a}}, {grizli}, 1.8.2, Zenodo,  Zenodo, \dodoi{10.5281/zenodo.7712834}

\bibitem[{{Brammer}(2023{\natexlab{b}})}]{griz2023}
---. 2023{\natexlab{b}}, {grizli}, 1.9.11, Zenodo,  Zenodo, \dodoi{10.5281/zenodo.1146904}

\bibitem[{{Brinchmann} {et~al.}(2004){Brinchmann}, {Charlot}, {Heckman}, {Kauffmann}, {Tremonti}, \& {White}}]{2004astro.ph..6220B}
{Brinchmann}, J., {Charlot}, S., {Heckman}, T.~M., {et~al.} 2004, arXiv e-prints, astro, \dodoi{10.48550/arXiv.astro-ph/0406220}

\bibitem[{Bushouse {et~al.}(2023)Bushouse, Eisenhamer, Dencheva, Davies, Greenfield, Morrison, Hodge, Simon, Grumm, Droettboom, Slavich, Sosey, Pauly, Miller, Jedrzejewski, Hack, Davis, Crawford, Law, Gordon, Regan, Cara, MacDonald, Bradley, Shanahan, Jamieson, Teodoro, Williams, \& Pena-Guerrero}]{calwebb_v1.12.5_2023}
Bushouse, H., Eisenhamer, J., Dencheva, N., {et~al.} 2023, JWST Calibration Pipeline, 1.12.5,  Zenodo, \dodoi{10.5281/zenodo.10022973}

\bibitem[{{Cameron} {et~al.}(2023){Cameron}, {Saxena}, {Bunker}, {D'Eugenio}, {Carniani}, {Maiolino}, {Curtis-Lake}, {Ferruit}, {Jakobsen}, {Arribas}, {Bonaventura}, {Charlot}, {Chevallard}, {Curti}, {Looser}, {Maseda}, {Rawle}, {Rodr{\'\i}guez Del Pino}, {Smit}, {{\"U}bler}, {Willott}, {Witstok}, {Egami}, {Eisenstein}, {Johnson}, {Hainline}, {Rieke}, {Robertson}, {Stark}, {Tacchella}, {Williams}, {Willmer}, {Bhatawdekar}, {Bowler}, {Boyett}, {Circosta}, {Helton}, {Jones}, {Kumari}, {Ji}, {Nelson}, {Parlanti}, {Sandles}, {Scholtz}, \& {Sun}}]{Cameron2023}
{Cameron}, A.~J., {Saxena}, A., {Bunker}, A.~J., {et~al.} 2023, \aap, 677, A115, \dodoi{10.1051/0004-6361/202346107}

\bibitem[{{Cardelli} {et~al.}(1989){Cardelli}, {Clayton}, \& {Mathis}}]{1989ApJ...345..245C}
{Cardelli}, J.~A., {Clayton}, G.~C., \& {Mathis}, J.~S. 1989, \apj, 345, 245, \dodoi{10.1086/167900}

\bibitem[{{Carnall} {et~al.}(2024){Carnall}, {Cullen}, {McLure}, {McLeod}, {Begley}, {Donnan}, {Dunlop}, {Shapley}, {Rowlands}, {Almaini}, {Arellano-C{\'o}rdova}, {Barrufet}, {Cimatti}, {Ellis}, {Grogin}, {Hamadouche}, {Illingworth}, {Koekemoer}, {Leung}, {Lovell}, {P{\'e}rez-Gonz{\'a}lez}, {Santini}, {Stanton}, \& {Wild}}]{Carnall2024}
{Carnall}, A.~C., {Cullen}, F., {McLure}, R.~J., {et~al.} 2024, \mnras, 534, 325, \dodoi{10.1093/mnras/stae2092}

\bibitem[{{Cataldi} {et~al.}(2025){Cataldi}, {Belfiore}, {Curti}, {Moreschini}, {Mannucci}, {D'Amato}, {Cresci}, {Feltre}, {Ginolfi}, {Marconi}, {Amiri}, {Arnaboldi}, {Bertola}, {Bracci}, {Carniani}, {Ceci}, {Chakraborty}, {Cirasuolo}, {Cullen}, {Kobayashi}, {Kumari}, {Maiolino}, {Marconcini}, {Scialpi}, \& {Ulivi}}]{Cataldi2025}
{Cataldi}, E., {Belfiore}, F., {Curti}, M., {et~al.} 2025, arXiv e-prints, arXiv:2504.03839, \dodoi{10.48550/arXiv.2504.03839}

\bibitem[{{Chabrier}(2003)}]{Chabrier2003}
{Chabrier}, G. 2003, \pasp, 115, 763, \dodoi{10.1086/376392}

\bibitem[{{Clarke} {et~al.}(2023){Clarke}, {Shapley}, {Sanders}, {Topping}, {Jones}, {Kriek}, {Reddy}, {Stark}, \& {Tang}}]{Clarke2023}
{Clarke}, L., {Shapley}, A., {Sanders}, R.~L., {et~al.} 2023, \apj, 957, 81, \dodoi{10.3847/1538-4357/acfedb}

\bibitem[{{Cullen} {et~al.}(2019){Cullen}, {McLure}, {Dunlop}, {Khochfar}, {Dav{\'e}}, {Amor{\'\i}n}, {Bolzonella}, {Carnall}, {Castellano}, {Cimatti}, {Cirasuolo}, {Cresci}, {Fynbo}, {Fontanot}, {Gargiulo}, {Garilli}, {Guaita}, {Hathi}, {Hibon}, {Mannucci}, {Marchi}, {McLeod}, {Pentericci}, {Pozzetti}, {Shapley}, {Talia}, \& {Zamorani}}]{cull2019}
{Cullen}, F., {McLure}, R.~J., {Dunlop}, J.~S., {et~al.} 2019, \mnras, 487, 2038, \dodoi{10.1093/mnras/stz1402}

\bibitem[{{Cullen} {et~al.}(2020){Cullen}, {McLure}, {Dunlop}, {Carnall}, {McLeod}, {Shapley}, {Amor{\'\i}n}, {Bolzonella}, {Castellano}, {Cimatti}, {Cirasuolo}, {Cucciati}, {Fontana}, {Fontanot}, {Garilli}, {Guaita}, {Jarvis}, {Pentericci}, {Pozzetti}, {Talia}, {Zamorani}, {Calabr{\`o}}, {Cresci}, {Fynbo}, {Hathi}, {Giavalisco}, {Koekemoer}, {Mannucci}, \& {Saxena}}]{Cullen2020}
---. 2020, \mnras, 495, 1501, \dodoi{10.1093/mnras/staa1260}

\bibitem[{{de Barros} {et~al.}(2016){de Barros}, {Vanzella}, {Amor{\'\i}n}, {Castellano}, {Siana}, {Grazian}, {Suh}, {Balestra}, {Vignali}, {Verhamme}, {Zamorani}, {Mignoli}, {Hasinger}, {Comastri}, {Pentericci}, {P{\'e}rez-Montero}, {Fontana}, {Giavalisco}, \& {Gilli}}]{deBarros2016}
{de Barros}, S., {Vanzella}, E., {Amor{\'\i}n}, R., {et~al.} 2016, \aap, 585, A51, \dodoi{10.1051/0004-6361/201527046}

\bibitem[{{Dijkstra} {et~al.}(2016){Dijkstra}, {Gronke}, \& {Venkatesan}}]{Dijkstra2016}
{Dijkstra}, M., {Gronke}, M., \& {Venkatesan}, A. 2016, \apj, 828, 71, \dodoi{10.3847/0004-637X/828/2/71}

\bibitem[{{Du} {et~al.}(2021){Du}, {Shapley}, {Topping}, {Reddy}, {Sanders}, {Coil}, {Kriek}, {Mobasher}, \& {Siana}}]{Du2021}
{Du}, X., {Shapley}, A.~E., {Topping}, M.~W., {et~al.} 2021, \apj, 920, 95, \dodoi{10.3847/1538-4357/ac1273}

\bibitem[{{Ferland} {et~al.}(2013){Ferland}, {Porter}, {van Hoof}, {Williams}, {Abel}, {Lykins}, {Shaw}, {Henney}, \& {Stancil}}]{ferl2013}
{Ferland}, G.~J., {Porter}, R.~L., {van Hoof}, P.~A.~M., {et~al.} 2013, \rmxaa, 49, 137.
\newblock \doarXiv{1302.4485}

\bibitem[{{Fletcher} {et~al.}(2019){Fletcher}, {Tang}, {Robertson}, {Nakajima}, {Ellis}, {Stark}, \& {Inoue}}]{Fletcher2019}
{Fletcher}, T.~J., {Tang}, M., {Robertson}, B.~E., {et~al.} 2019, \apj, 878, 87, \dodoi{10.3847/1538-4357/ab2045}

\bibitem[{{Garg} {et~al.}(2022){Garg}, {Narayanan}, {Byler}, {Sanders}, {Shapley}, {Strom}, {Dav{\'e}}, {Hirschmann}, {Lovell}, {Otter}, {Popping}, \& {Privon}}]{Garg2022}
{Garg}, P., {Narayanan}, D., {Byler}, N., {et~al.} 2022, \apj, 926, 80, \dodoi{10.3847/1538-4357/ac43b8}

\bibitem[{{Gburek} {et~al.}(2019){Gburek}, {Siana}, {Alavi}, {Emami}, {Richard}, {Freeman}, {Stark}, {Snapp-Kolas}, \& {Lucero}}]{Gburek2019}
{Gburek}, T., {Siana}, B., {Alavi}, A., {et~al.} 2019, \apj, 887, 168, \dodoi{10.3847/1538-4357/ab5713}

\bibitem[{{Guaita} {et~al.}(2011){Guaita}, {Acquaviva}, {Padilla}, {Gawiser}, {Bond}, {Ciardullo}, {Treister}, {Kurczynski}, {Gronwall}, {Lira}, \& {Schawinski}}]{Guaita2011}
{Guaita}, L., {Acquaviva}, V., {Padilla}, N., {et~al.} 2011, \apj, 733, 114, \dodoi{10.1088/0004-637X/733/2/114}

\bibitem[{{Harris} {et~al.}(2020){Harris}, {Millman}, {van der Walt}, {Gommers}, {Virtanen}, {Cournapeau}, {Wieser}, {Taylor}, {Berg}, {Smith}, {Kern}, {Picus}, {Hoyer}, {van Kerkwijk}, {Brett}, {Haldane}, {del R{\'\i}o}, {Wiebe}, {Peterson}, {G{\'e}rard-Marchant}, {Sheppard}, {Reddy}, {Weckesser}, {Abbasi}, {Gohlke}, \& {Oliphant}}]{harr2020}
{Harris}, C.~R., {Millman}, K.~J., {van der Walt}, S.~J., {et~al.} 2020, \nat, 585, 357, \dodoi{10.1038/s41586-020-2649-2}

\bibitem[{{Horne}(1986)}]{Horne1986}
{Horne}, K. 1986, \pasp, 98, 609, \dodoi{10.1086/131801}

\bibitem[{{Hu} {et~al.}(2017){Hu}, {Wang}, {Zheng}, {Malhotra}, {Infante}, {Rhoads}, {Gonzalez}, {Walker}, {Jiang}, {Jiang}, {Hibon}, {Barrientos}, {Finkelstein}, {Galaz}, {Kang}, {Kong}, {Tilvi}, {Yang}, \& {Zheng}}]{Hu2017}
{Hu}, W., {Wang}, J., {Zheng}, Z.-Y., {et~al.} 2017, \apjl, 845, L16, \dodoi{10.3847/2041-8213/aa8401}

\bibitem[{{Hunter}(2007)}]{hunt2007}
{Hunter}, J.~D. 2007, Computing in Science and Engineering, 9, 90, \dodoi{10.1109/MCSE.2007.55}

\bibitem[{{Ilyushin}(2024)}]{Ilyushin2024}
{Ilyushin}, B.~B. 2024, Journal of Engineering Thermophysics, 33, 1, \dodoi{10.1134/S1810232824010016}

\bibitem[{{Izotov} {et~al.}(2016){Izotov}, {Orlitov{\'a}}, {Schaerer}, {Thuan}, {Verhamme}, {Guseva}, \& {Worseck}}]{izot2016a}
{Izotov}, Y.~I., {Orlitov{\'a}}, I., {Schaerer}, D., {et~al.} 2016, \nat, 529, 178, \dodoi{10.1038/nature16456}

\bibitem[{{Izotov} {et~al.}(2018){Izotov}, {Schaerer}, {Worseck}, {Guseva}, {Thuan}, {Verhamme}, {Orlitov{\'a}}, \& {Fricke}}]{izot2018a}
{Izotov}, Y.~I., {Schaerer}, D., {Worseck}, G., {et~al.} 2018, \mnras, 474, 4514, \dodoi{10.1093/mnras/stx3115}

\bibitem[{{Kauffmann} {et~al.}(2003){Kauffmann}, {Heckman}, {White}, {Charlot}, {Tremonti}, {Brinchmann}, {Bruzual}, {Peng}, {Seibert}, {Bernardi}, {Blanton}, {Brinkmann}, {Castander}, {Cs{\'a}bai}, {Fukugita}, {Ivezic}, {Munn}, {Nichol}, {Padmanabhan}, {Thakar}, {Weinberg}, \& {York}}]{Kauffmann2003}
{Kauffmann}, G., {Heckman}, T.~M., {White}, S. D.~M., {et~al.} 2003, \mnras, 341, 33, \dodoi{10.1046/j.1365-8711.2003.06291.x}

\bibitem[{{Kennicutt}(1998)}]{Kennicutt1998}
{Kennicutt}, Robert~C., J. 1998, \araa, 36, 189, \dodoi{10.1146/annurev.astro.36.1.189}

\bibitem[{{Kewley} {et~al.}(2001){Kewley}, {Dopita}, {Sutherland}, {Heisler}, \& {Trevena}}]{Kewley2001}
{Kewley}, L.~J., {Dopita}, M.~A., {Sutherland}, R.~S., {Heisler}, C.~A., \& {Trevena}, J. 2001, \apj, 556, 121, \dodoi{10.1086/321545}

\bibitem[{{Kewley} {et~al.}(2006){Kewley}, {Groves}, {Kauffmann}, \& {Heckman}}]{Kewley2006}
{Kewley}, L.~J., {Groves}, B., {Kauffmann}, G., \& {Heckman}, T. 2006, \mnras, 372, 961, \dodoi{10.1111/j.1365-2966.2006.10859.x}

\bibitem[{{Kewley} {et~al.}(2019){Kewley}, {Nicholls}, \& {Sutherland}}]{kewley2019}
{Kewley}, L.~J., {Nicholls}, D.~C., \& {Sutherland}, R.~S. 2019, \araa, 57, 511, \dodoi{10.1146/annurev-astro-081817-051832}

\bibitem[{{Korhonen Cuestas} {et~al.}(2025){Korhonen Cuestas}, {Strom}, {Miller}, {Steidel}, {Trainor}, {Rudie}, \& {Nu{\~n}ez}}]{korh2025}
{Korhonen Cuestas}, N.~A., {Strom}, A.~L., {Miller}, T.~B., {et~al.} 2025, arXiv e-prints, arXiv:2503.10800, \dodoi{10.48550/arXiv.2503.10800}

\bibitem[{{Laigle} {et~al.}(2016){Laigle}, {McCracken}, {Ilbert}, {Hsieh}, {Davidzon}, {Capak}, {Hasinger}, {Silverman}, {Pichon}, {Coupon}, {Aussel}, {Le Borgne}, {Caputi}, {Cassata}, {Chang}, {Civano}, {Dunlop}, {Fynbo}, {Kartaltepe}, {Koekemoer}, {Le F{\`e}vre}, {Le Floc'h}, {Leauthaud}, {Lilly}, {Lin}, {Marchesi}, {Milvang-Jensen}, {Salvato}, {Sanders}, {Scoville}, {Smolcic}, {Stockmann}, {Taniguchi}, {Tasca}, {Toft}, {Vaccari}, \& {Zabl}}]{Laigle2016}
{Laigle}, C., {McCracken}, H.~J., {Ilbert}, O., {et~al.} 2016, \apjs, 224, 24, \dodoi{10.3847/0067-0049/224/2/24}

\bibitem[{{Law} {et~al.}(2021){Law}, {Ji}, {Belfiore}, {Bershady}, {Cappellari}, {Westfall}, {Yan}, {Bizyaev}, {Brownstein}, {Drory}, \& {Andrews}}]{Law2021}
{Law}, D.~R., {Ji}, X., {Belfiore}, F., {et~al.} 2021, \apj, 915, 35, \dodoi{10.3847/1538-4357/abfe0a}

\bibitem[{{Madau} \& {Dickinson}(2014)}]{2014ARA&A..52..415M}
{Madau}, P., \& {Dickinson}, M. 2014, \araa, 52, 415, \dodoi{10.1146/annurev-astro-081811-125615}

\bibitem[{{Maseda} {et~al.}(2023){Maseda}, {Lewis}, {Matthee}, {Hennawi}, {Boogaard}, {Feltre}, {Nanayakkara}, {Bacon}, {Barger}, {Brinchmann}, {Franx}, {Hashimoto}, {Inami}, {Kusakabe}, {Leclercq}, {Rowland}, {Taylor}, {Tremonti}, {Urrutia}, {Schaye}, {Simmonds}, \& {Vitte}}]{Maseda2023}
{Maseda}, M.~V., {Lewis}, Z., {Matthee}, J., {et~al.} 2023, \apj, 956, 11, \dodoi{10.3847/1538-4357/acf12b}

\bibitem[{{Masters} {et~al.}(2016){Masters}, {Faisst}, \& {Capak}}]{masters2016}
{Masters}, D., {Faisst}, A., \& {Capak}, P. 2016, \apj, 828, 18, \dodoi{10.3847/0004-637X/828/1/18}

\bibitem[{{Matthee} \& {Schaye}(2018)}]{matthee2018}
{Matthee}, J., \& {Schaye}, J. 2018, \mnras, 479, L34, \dodoi{10.1093/mnrasl/sly093}

\bibitem[{{Matthee} {et~al.}(2021){Matthee}, {Sobral}, {Hayes}, {Pezzulli}, {Gronke}, {Schaerer}, {Naidu}, {R{\"o}ttgering}, {Calhau}, {Paulino-Afonso}, {Santos}, \& {Amor{\'\i}n}}]{Matthee2021}
{Matthee}, J., {Sobral}, D., {Hayes}, M., {et~al.} 2021, \mnras, 505, 1382, \dodoi{10.1093/mnras/stab1304}

\bibitem[{{McLean} {et~al.}(2012){McLean}, {Steidel}, {Epps}, {Konidaris}, {Matthews}, {Adkins}, {Aliado}, {Brims}, {Canfield}, {Cromer}, {Fucik}, {Kulas}, {Mace}, {Magnone}, {Rodriguez}, {Rudie}, {Trainor}, {Wang}, {Weber}, \& {Weiss}}]{mclean2012}
{McLean}, I.~S., {Steidel}, C.~C., {Epps}, H.~W., {et~al.} 2012, in Society of Photo-Optical Instrumentation Engineers (SPIE) Conference Series, Vol. 8446, Ground-based and Airborne Instrumentation for Astronomy IV, ed. I.~S. {McLean}, S.~K. {Ramsay}, \& H.~{Takami}, 84460J, \dodoi{10.1117/12.924794}

\bibitem[{{Messa} {et~al.}(2025){Messa}, {Vanzella}, {Loiacono}, {Bergamini}, {Castellano}, {Sun}, {Willott}, {Windhorst}, {Yan}, {Angora}, {Rosati}, {Adamo}, {Annibali}, {Bolamperti}, {Brada{\v{c}}}, {Bradley}, {Calura}, {Claeyssens}, {Comastri}, {Conselice}, {D'Silva}, {Dickinson}, {Frye}, {Grillo}, {Grogin}, {Gruppioni}, {Koekemoer}, {Meneghetti}, {Me{\v{s}}tri{\'c}}, {Pascale}, {Ravindranath}, {Ricotti}, {Summers}, \& {Zanella}}]{Messa25}
{Messa}, M., {Vanzella}, E., {Loiacono}, F., {et~al.} 2025, \aap, 694, A59, \dodoi{10.1051/0004-6361/202451695}

\bibitem[{{Nakajima} {et~al.}(2012){Nakajima}, {Ouchi}, {Shimasaku}, {Ono}, {Lee}, {Foucaud}, {Ly}, {Dale}, {Salim}, {Finn}, {Almaini}, \& {Okamura}}]{Nakajima2012}
{Nakajima}, K., {Ouchi}, M., {Shimasaku}, K., {et~al.} 2012, \apj, 745, 12, \dodoi{10.1088/0004-637X/745/1/12}

\bibitem[{{Oke} {et~al.}(1995){Oke}, {Cohen}, {Carr}, {Cromer}, {Dingizian}, {Harris}, {Labrecque}, {Lucinio}, {Schaal}, {Epps}, \& {Miller}}]{oke1995}
{Oke}, J.~B., {Cohen}, J.~G., {Carr}, M., {et~al.} 1995, \pasp, 107, 375, \dodoi{10.1086/133562}

\bibitem[{{Osterbrock} \& {Ferland}(2006)}]{Osterbrock2006}
{Osterbrock}, D.~E., \& {Ferland}, G.~J. 2006, {Astrophysics of gaseous nebulae and active galactic nuclei}

\bibitem[{{{\"O}stlin} {et~al.}(2014){{\"O}stlin}, {Hayes}, {Duval}, {Sandberg}, {Rivera-Thorsen}, {Marquart}, {Orlitov{\'a}}, {Adamo}, {Melinder}, {Guaita}, {Atek}, {Cannon}, {Gruyters}, {Herenz}, {Kunth}, {Laursen}, {Mas-Hesse}, {Micheva}, {Ot{\'\i}-Floranes}, {Pardy}, {Roth}, {Schaerer}, \& {Verhamme}}]{Ostlin2014}
{{\"O}stlin}, G., {Hayes}, M., {Duval}, F., {et~al.} 2014, \apj, 797, 11, \dodoi{10.1088/0004-637X/797/1/11}

\bibitem[{{Oteo} {et~al.}(2015){Oteo}, {Sobral}, {Ivison}, {Smail}, {Best}, {Cepa}, \& {P{\'e}rez-Garc{\'\i}a}}]{Oteo2015}
{Oteo}, I., {Sobral}, D., {Ivison}, R.~J., {et~al.} 2015, \mnras, 452, 2018, \dodoi{10.1093/mnras/stv1284}

\bibitem[{{Pentericci} {et~al.}(2011){Pentericci}, {Fontana}, {Vanzella}, {Castellano}, {Grazian}, {Dijkstra}, {Boutsia}, {Cristiani}, {Dickinson}, {Giallongo}, {Giavalisco}, {Maiolino}, {Moorwood}, {Paris}, \& {Santini}}]{Pentericci2011}
{Pentericci}, L., {Fontana}, A., {Vanzella}, E., {et~al.} 2011, \apj, 743, 132, \dodoi{10.1088/0004-637X/743/2/132}

\bibitem[{{Rauscher}(2023)}]{raus2023}
{Rauscher}, B.~J. 2023, arXiv e-prints, arXiv:2306.03250, \dodoi{10.48550/arXiv.2306.03250}

\bibitem[{{Reddy} \& {Steidel}(2009)}]{reddy2009}
{Reddy}, N.~A., \& {Steidel}, C.~C. 2009, \apj, 692, 778, \dodoi{10.1088/0004-637X/692/1/778}

\bibitem[{{Reddy} {et~al.}(2020){Reddy}, {Shapley}, {Kriek}, {Steidel}, {Shivaei}, {Sanders}, {Mobasher}, {Coil}, {Siana}, {Freeman}, {Azadi}, {Fetherolf}, {Leung}, {Price}, \& {Zick}}]{redd2020}
{Reddy}, N.~A., {Shapley}, A.~E., {Kriek}, M., {et~al.} 2020, \apj, 902, 123, \dodoi{10.3847/1538-4357/abb674}

\bibitem[{{Rogers} {et~al.}(2024){Rogers}, {Strom}, {Rudie}, {Trainor}, {Raptis}, \& {von Raesfeld}}]{2024ApJ...964L..12R}
{Rogers}, N. S.~J., {Strom}, A.~L., {Rudie}, G.~C., {et~al.} 2024, \apjl, 964, L12, \dodoi{10.3847/2041-8213/ad2f37}

\bibitem[{{Rudie} {et~al.}(2012){Rudie}, {Steidel}, {Trainor}, {Rakic}, {Bogosavljevi{\'c}}, {Pettini}, {Reddy}, {Shapley}, {Erb}, \& {Law}}]{rudi2012}
{Rudie}, G.~C., {Steidel}, C.~C., {Trainor}, R.~F., {et~al.} 2012, \apj, 750, 67, \dodoi{10.1088/0004-637X/750/1/67}

\bibitem[{{Runnholm} {et~al.}(2020){Runnholm}, {Hayes}, {Melinder}, {Rivera-Thorsen}, {{\"O}stlin}, {Cannon}, \& {Kunth}}]{Runnholm2020}
{Runnholm}, A., {Hayes}, M., {Melinder}, J., {et~al.} 2020, \apj, 892, 48, \dodoi{10.3847/1538-4357/ab7a91}

\bibitem[{{Runnholm} {et~al.}(2025){Runnholm}, {Hayes}, {Mehta}, {Malkan}, {Scarlata}, {Nedkova}, {Rafelski}, {Vulcani}, {Huberty}, {Herenz}, {Hutter}, {Bruton}, {Acharyya}, {Atek}, {Baronchelli}, {Battisti}, {Brada{\v{c}}}, {Bunker}, {Dai}, {Hannahs}, {Hasan}, {Kim}, {Leethochawalit}, {Lin}, {Rutkowski}, {Saldana-Lopez}, {Sattari}, \& {Wang}}]{Runnholm25}
{Runnholm}, A., {Hayes}, M.~J., {Mehta}, V., {et~al.} 2025, arXiv e-prints, arXiv:2502.19174, \dodoi{10.48550/arXiv.2502.19174}

\bibitem[{{Sanders} {et~al.}(2023){Sanders}, {Shapley}, {Topping}, {Reddy}, \& {Brammer}}]{Sanders2023}
{Sanders}, R.~L., {Shapley}, A.~E., {Topping}, M.~W., {Reddy}, N.~A., \& {Brammer}, G.~B. 2023, \apj, 955, 54, \dodoi{10.3847/1538-4357/acedad}

\bibitem[{{Sanders} {et~al.}(2024){Sanders}, {Shapley}, {Topping}, {Reddy}, \& {Brammer}}]{Sanders2024}
---. 2024, \apj, 962, 24, \dodoi{10.3847/1538-4357/ad15fc}

\bibitem[{{Sanders} {et~al.}(2016){Sanders}, {Shapley}, {Kriek}, {Reddy}, {Freeman}, {Coil}, {Siana}, {Mobasher}, {Shivaei}, {Price}, \& {de Groot}}]{sand2016}
{Sanders}, R.~L., {Shapley}, A.~E., {Kriek}, M., {et~al.} 2016, \apj, 816, 23, \dodoi{10.3847/0004-637X/816/1/23}

\bibitem[{{Sanders} {et~al.}(2020){Sanders}, {Shapley}, {Reddy}, {Kriek}, {Siana}, {Coil}, {Mobasher}, {Shivaei}, {Freeman}, {Azadi}, {Price}, {Leung}, {Fetherolf}, {de Groot}, {Zick}, {Fornasini}, \& {Barro}}]{sand2020}
{Sanders}, R.~L., {Shapley}, A.~E., {Reddy}, N.~A., {et~al.} 2020, \mnras, 491, 1427, \dodoi{10.1093/mnras/stz3032}

\bibitem[{{Sanders} {et~al.}(2021){Sanders}, {Shapley}, {Jones}, {Reddy}, {Kriek}, {Siana}, {Coil}, {Mobasher}, {Shivaei}, {Dav{\'e}}, {Azadi}, {Price}, {Leung}, {Freeman}, {Fetherolf}, {de Groot}, {Zick}, \& {Barro}}]{Sanders2021}
{Sanders}, R.~L., {Shapley}, A.~E., {Jones}, T., {et~al.} 2021, \apj, 914, 19, \dodoi{10.3847/1538-4357/abf4c1}

\bibitem[{{Schenker} {et~al.}(2012){Schenker}, {Stark}, {Ellis}, {Robertson}, {Dunlop}, {McLure}, {Kneib}, \& {Richard}}]{Schenker2012}
{Schenker}, M.~A., {Stark}, D.~P., {Ellis}, R.~S., {et~al.} 2012, \apj, 744, 179, \dodoi{10.1088/0004-637X/744/2/179}

\bibitem[{{Scholtz} {et~al.}(2023){Scholtz}, {Maiolino}, {D'Eugenio}, {Curtis-Lake}, {Carniani}, {Charlot}, {Curti}, {Silcock}, {Arribas}, {Baker}, {Bhatawdekar}, {Boyett}, {Bunker}, {Chevallard}, {Circosta}, {Eisenstein}, {Hainline}, {Hausen}, {Ji}, {Ji}, {Johnson}, {Kumari}, {Looser}, {Lyu}, {Maseda}, {Parlanti}, {Perna}, {Rieke}, {Robertson}, {Rodr{\'\i}guez Del Pino}, {Sun}, {Tacchella}, {{\"U}bler}, {Venturi}, {Williams}, {Willmer}, {Willott}, \& {Witstok}}]{Scholtz2023}
{Scholtz}, J., {Maiolino}, R., {D'Eugenio}, F., {et~al.} 2023, arXiv e-prints, arXiv:2311.18731, \dodoi{10.48550/arXiv.2311.18731}

\bibitem[{{Shapley} {et~al.}(2015){Shapley}, {Reddy}, {Kriek}, {Freeman}, {Sanders}, {Siana}, {Coil}, {Mobasher}, {Shivaei}, {Price}, \& {de Groot}}]{Shapley2015}
{Shapley}, A.~E., {Reddy}, N.~A., {Kriek}, M., {et~al.} 2015, \apj, 801, 88, \dodoi{10.1088/0004-637X/801/2/88}

\bibitem[{{Shapley} {et~al.}(2019){Shapley}, {Sanders}, {Shao}, {Reddy}, {Kriek}, {Coil}, {Mobasher}, {Siana}, {Shivaei}, {Freeman}, {Azadi}, {Price}, {Leung}, {Fetherolf}, {de Groot}, {Zick}, {Fornasini}, \& {Barro}}]{Shapley2019}
{Shapley}, A.~E., {Sanders}, R.~L., {Shao}, P., {et~al.} 2019, \apjl, 881, L35, \dodoi{10.3847/2041-8213/ab385a}

\bibitem[{{Shapley} {et~al.}(2021){Shapley}, {Sanders}, {Berg}, {Bouwens}, {Brammer}, {Cullen}, {Dave}, {Du}, {Dunlop}, {Ellis}, {Forster Schreiber}, {Furlanetto}, {Glazebrook}, {Illingworth}, {Jones}, {Kriek}, {McLure}, {Narayanan}, {Oesch}, {Pahl}, {Pettini}, {Reddy}, {Runco}, {Schaerer}, {Stark}, {Steidel}, {Tang}, \& {Topping}}]{Shapley21}
{Shapley}, A.~E., {Sanders}, R., {Berg}, D., {et~al.} 2021, {The AURORA Survey: First Direct Metallicity Calibrations at High Redshift}, JWST Proposal. Cycle 1, ID. \#1914

\bibitem[{{Stanway} \& {Eldridge}(2018)}]{Stanway2018}
{Stanway}, E.~R., \& {Eldridge}, J.~J. 2018, \mnras, 479, 75, \dodoi{10.1093/mnras/sty1353}

\bibitem[{{Stark} {et~al.}(2010){Stark}, {Ellis}, {Chiu}, {Ouchi}, \& {Bunker}}]{Stark2010}
{Stark}, D.~P., {Ellis}, R.~S., {Chiu}, K., {Ouchi}, M., \& {Bunker}, A. 2010, \mnras, 408, 1628, \dodoi{10.1111/j.1365-2966.2010.17227.x}

\bibitem[{{Stark} {et~al.}(2011){Stark}, {Ellis}, \& {Ouchi}}]{Stark2011}
{Stark}, D.~P., {Ellis}, R.~S., \& {Ouchi}, M. 2011, \apjl, 728, L2, \dodoi{10.1088/2041-8205/728/1/L2}

\bibitem[{{Steidel} {et~al.}(2003){Steidel}, {Adelberger}, {Shapley}, {Pettini}, {Dickinson}, \& {Giavalisco}}]{steidel2003}
{Steidel}, C.~C., {Adelberger}, K.~L., {Shapley}, A.~E., {et~al.} 2003, \apj, 592, 728, \dodoi{10.1086/375772}

\bibitem[{{Steidel} {et~al.}(2018){Steidel}, {Bogosavljevi{\'c}}, {Shapley}, {Reddy}, {Rudie}, {Pettini}, {Trainor}, \& {Strom}}]{stei2018}
{Steidel}, C.~C., {Bogosavljevi{\'c}}, M., {Shapley}, A.~E., {et~al.} 2018, \apj, 869, 123, \dodoi{10.3847/1538-4357/aaed28}

\bibitem[{{Steidel} {et~al.}(2004){Steidel}, {Shapley}, {Pettini}, {Adelberger}, {Erb}, {Reddy}, \& {Hunt}}]{steidel2004}
{Steidel}, C.~C., {Shapley}, A.~E., {Pettini}, M., {et~al.} 2004, \apj, 604, 534, \dodoi{10.1086/381960}

\bibitem[{{Steidel} {et~al.}(2016){Steidel}, {Strom}, {Pettini}, {Rudie}, {Reddy}, \& {Trainor}}]{stei2016}
{Steidel}, C.~C., {Strom}, A.~L., {Pettini}, M., {et~al.} 2016, \apj, 826, 159, \dodoi{10.3847/0004-637X/826/2/159}

\bibitem[{{Steidel} {et~al.}(2014){Steidel}, {Rudie}, {Strom}, {Pettini}, {Reddy}, {Shapley}, {Trainor}, {Erb}, {Turner}, {Konidaris}, {Kulas}, {Mace}, {Matthews}, \& {McLean}}]{stei2014}
{Steidel}, C.~C., {Rudie}, G.~C., {Strom}, A.~L., {et~al.} 2014, \apj, 795, 165, \dodoi{10.1088/0004-637X/795/2/165}

\bibitem[{{Strom} {et~al.}(2022){Strom}, {Rudie}, {Steidel}, \& {Trainor}}]{Strom2022}
{Strom}, A.~L., {Rudie}, G.~C., {Steidel}, C.~C., \& {Trainor}, R.~F. 2022, \apj, 925, 116, \dodoi{10.3847/1538-4357/ac38a3}

\bibitem[{{Strom} {et~al.}(2018){Strom}, {Steidel}, {Rudie}, {Trainor}, \& {Pettini}}]{stro2018}
{Strom}, A.~L., {Steidel}, C.~C., {Rudie}, G.~C., {Trainor}, R.~F., \& {Pettini}, M. 2018, \apj, 868, 117, \dodoi{10.3847/1538-4357/aae1a5}

\bibitem[{{Strom} {et~al.}(2017){Strom}, {Steidel}, {Rudie}, {Trainor}, {Pettini}, \& {Reddy}}]{stro2017}
{Strom}, A.~L., {Steidel}, C.~C., {Rudie}, G.~C., {et~al.} 2017, \apj, 836, 164, \dodoi{10.3847/1538-4357/836/2/164}

\bibitem[{{Strom} {et~al.}(2023){Strom}, {Rudie}, {Trainor}, {Brammer}, {Maseda}, {Raptis}, {Rogers}, {Steidel}, {Chen}, \& {Law}}]{stro2023}
{Strom}, A.~L., {Rudie}, G.~C., {Trainor}, R.~F., {et~al.} 2023, \apjl, 958, L11, \dodoi{10.3847/2041-8213/ad07dc}

\bibitem[{{Sutherland} \& {Dopita}(2017)}]{Sutherland2017}
{Sutherland}, R.~S., \& {Dopita}, M.~A. 2017, \apjs, 229, 34, \dodoi{10.3847/1538-4365/aa6541}

\bibitem[{{Tang} {et~al.}(2019){Tang}, {Stark}, {Chevallard}, \& {Charlot}}]{Tang2019}
{Tang}, M., {Stark}, D.~P., {Chevallard}, J., \& {Charlot}, S. 2019, \mnras, 489, 2572, \dodoi{10.1093/mnras/stz2236}

\bibitem[{{Topping} {et~al.}(2020){Topping}, {Shapley}, {Reddy}, {Sanders}, {Coil}, {Kriek}, {Mobasher}, \& {Siana}}]{topp2020}
{Topping}, M.~W., {Shapley}, A.~E., {Reddy}, N.~A., {et~al.} 2020, \mnras, 499, 1652, \dodoi{10.1093/mnras/staa2941}

\bibitem[{{Trainor} {et~al.}(2025){Trainor}, {Lamb}, {Steidel}, {Chen}, {Erb}, {Trenholm}, {McClain}, \& {Kovach}}]{Trainor2025}
{Trainor}, R.~F., {Lamb}, N.~R., {Steidel}, C.~C., {et~al.} 2025, arXiv e-prints, arXiv:2505.15881, \dodoi{10.48550/arXiv.2505.15881}

\bibitem[{{Trainor} {et~al.}(2015){Trainor}, {Steidel}, {Strom}, \& {Rudie}}]{trai2015}
{Trainor}, R.~F., {Steidel}, C.~C., {Strom}, A.~L., \& {Rudie}, G.~C. 2015, \apj, 809, 89, \dodoi{10.1088/0004-637X/809/1/89}

\bibitem[{{Trainor} {et~al.}(2016){Trainor}, {Strom}, {Steidel}, \& {Rudie}}]{Trainor2016}
{Trainor}, R.~F., {Strom}, A.~L., {Steidel}, C.~C., \& {Rudie}, G.~C. 2016, \apj, 832, 171, \dodoi{10.3847/0004-637X/832/2/171}

\bibitem[{{Trainor} {et~al.}(2019){Trainor}, {Strom}, {Steidel}, {Rudie}, {Chen}, \& {Theios}}]{Trainor2019}
{Trainor}, R.~F., {Strom}, A.~L., {Steidel}, C.~C., {et~al.} 2019, \apj, 887, 85, \dodoi{10.3847/1538-4357/ab4993}

\bibitem[{{Tremonti} {et~al.}(2004){Tremonti}, {Heckman}, {Kauffmann}, {Brinchmann}, {Charlot}, {White}, {Seibert}, {Peng}, {Schlegel}, {Uomoto}, {Fukugita}, \& {Brinkmann}}]{trem2004}
{Tremonti}, C.~A., {Heckman}, T.~M., {Kauffmann}, G., {et~al.} 2004, \apj, 613, 898, \dodoi{10.1086/423264}

\bibitem[{{Veilleux} \& {Osterbrock}(1987)}]{Veilleux1987}
{Veilleux}, S., \& {Osterbrock}, D.~E. 1987, \apjs, 63, 295, \dodoi{10.1086/191166}

\bibitem[{{Verhamme} {et~al.}(2015){Verhamme}, {Orlitov{\'a}}, {Schaerer}, \& {Hayes}}]{verh2015}
{Verhamme}, A., {Orlitov{\'a}}, I., {Schaerer}, D., \& {Hayes}, M. 2015, \aap, 578, A7, \dodoi{10.1051/0004-6361/201423978}

\bibitem[{{Virtanen} {et~al.}(2020){Virtanen}, {Gommers}, {Oliphant}, {Haberland}, {Reddy}, {Cournapeau}, {Burovski}, {Peterson}, {Weckesser}, {Bright}, {van der Walt}, {Brett}, {Wilson}, {Millman}, {Mayorov}, {Nelson}, {Jones}, {Kern}, {Larson}, {Carey}, {Polat}, {Feng}, {Moore}, {VanderPlas}, {Laxalde}, {Perktold}, {Cimrman}, {Henriksen}, {Quintero}, {Harris}, {Archibald}, {Ribeiro}, {Pedregosa}, {van Mulbregt}, \& {SciPy 1. 0 Contributors}}]{virt2020}
{Virtanen}, P., {Gommers}, R., {Oliphant}, T.~E., {et~al.} 2020, Nature Methods, 17, 261, \dodoi{10.1038/s41592-019-0686-2}

\bibitem[{{Wang} {et~al.}(2018){Wang}, {Luo}, {Shen}, {Hou}, {Kong}, {Song}, {Zhang}, {Wu}, {Cao}, {Hou}, {Wang}, {Zhang}, \& {Zhao}}]{Wang2018}
{Wang}, L.-L., {Luo}, A.~L., {Shen}, S.-Y., {et~al.} 2018, \mnras, 474, 1873, \dodoi{10.1093/mnras/stx2798}

\bibitem[{{Willott} {et~al.}(2025){Willott}, {Asada}, {Iyer}, {Judez}, {Rihtarsic}, {Martis}, {Sarrouh}, {Desprez}, {Harshan}, {Mowla}, {Noirot}, {Felicioni}, {Bradac}, {Brammer}, {Muzzin}, {Sawicki}, {Antwi-Danso}, {Markov}, \& {Tripodi}}]{Willott25}
{Willott}, C.~J., {Asada}, Y., {Iyer}, K.~G., {et~al.} 2025, arXiv e-prints, arXiv:2502.07733, \dodoi{10.48550/arXiv.2502.07733}

\end{thebibliography}

\end{document}